\documentclass[aps,prd,10pt,groupedaddress,nofootinbib,amssymb,notitlepage,eqsecnum]{revtex4-1}
\usepackage{here}
\usepackage{graphicx}
\usepackage{amsmath}
\usepackage{bm}
\usepackage{color}
\usepackage[dvipsnames]{xcolor}
\usepackage[utf8]{inputenc}
\usepackage{amsfonts}
\definecolor{refs}{RGB}{245,156,74}
\usepackage[colorlinks=true,hyperfootnotes=true,citecolor=cyan]{hyperref}

\usepackage{amsfonts}
\usepackage{dcolumn}
\usepackage{hyperref}
\usepackage{stackengine}
\usepackage{enumitem}
\usepackage{comment}

\newcommand{\be}{\begin{equation}}  
\newcommand{\ee}{\end{equation}}
\newcommand{\ba}{\begin{eqnarray}}
\newcommand{\ea}{\end{eqnarray}}

\newcommand{\cH}{{\cal H}}

\newcommand{\bem}{\begin{bmatrix}}
\newcommand{\eem}{\end{bmatrix}}
\newcommand{\Mpl}{M_{\rm Pl}}

\allowdisplaybreaks

\begin{document}

\preprint{WUAP-26-?}

\title{Revisiting observational constraints on coupled exponential quintessence with \\
energy and momentum transfers: degeneracy with massive neutrinos}

\author{Jose Beltr\'an Jim\'enez$^{1,2,3}$, Kiyotomo Ichiki$^{4,5,6}$, Xiaolin Liu$^{7,6,8}$, Florencia Anabella Teppa Pannia$^{9,2,3}$, 
and Shinji Tsujikawa$^{10}$}

\affiliation{
$^{1}${\it Departamento de F\'isica Fundamental, Universidad de Salamanca, E-37008 Salamanca, Spain.}\\
$^{2}$ {\it Instituto Universitario de F\'isica Fundamental y Matem\'aticas~(IUFFyM), Universidad de Salamanca, E-37008 Salamanca, Spain.} \\
$^{3}$ {\it Institute of Theoretical Astrophysics, University of Oslo, N-0315 Oslo, Norway. }\\
$^{4}${\it Kobayashi-Maskawa Institute for the Origin of Particles and the Universe, Nagoya University, Furocho, Chikusa-ku, Nagoya, Aichi 464-8602, Japan.} \\
$^{5}${\it Institute for Advanced Research, Nagoya University, Furocho, Chikusa-ku, Nagoya, Aichi 464-8602, Japan.} \\
$^{6}${\it Graduate School of Science, Division of Particle and Astrophysical Science, Nagoya University, Furocho, Chikusa-ku, Nagoya, Aichi 464-8602, Japan.} \\
$^{7}${\it Department of Astronomy, Beijing Normal University,
Beijing 100875, China.} \\
$^{8}${\it Institudo de F\'isica Te\'orica UAM-CSIC, Universidad Aut\'onoma de Madrid, 28049, Spain.} \\
$^{9}$ {\it Departamento de Matem{\'a}tica Aplicada a la Ingenier{\'i}a Industrial, Universidad  Polit{\'e}cnica de Madrid, E-28006 Madrid, Spain.} \\
$^{10}${\it Department of Physics, Waseda University, 3-4-1 Okubo, Shinjuku, Tokyo 169-8555, Japan.}
}

\begin{abstract}
We investigate the impact of massive neutrinos on cosmological models in which dark energy, described by a quintessence scalar field $\phi$ with an exponential potential, interacts with dark matter through both energy and momentum transfers. Previous analyses have shown that the inclusion of low-redshift data tends to favour the detection of a pure momentum transfer between the dark sectors, consistent with the fact that such a transfer generically suppresses the growth of cosmic structures. 
Since massive neutrinos also reduce matter clustering, a potential degeneracy 
between the interaction parameters and the neutrino mass may arise. 
After updating the observational constraints on the model parameters obtained in earlier studies, we investigate the effect of allowing the neutrino mass to vary. We find that the detection of momentum transfer degrades once massive neutrinos are included. This occurs because a new degeneracy emerges between the neutrino mass and the parameter governing the energy exchange between dark energy and dark matter. Our findings differ from previous results in the literature, where the detection of momentum transfer was reported to be robust against varying neutrino masses. 
This suggests that the robustness of such detections depends on the underlying model and should therefore be carefully reassessed for each specific interacting scenario.

\end{abstract}

\date{\today}

\maketitle

\section{Introduction}
\label{introsec}

Understanding the nature of the dark components of the Universe remains one of the most pressing challenges in modern cosmology \cite{Sahni:1999gb,Padmanabhan:2002ji,Peebles:2002gy,Copeland:2006wr,Silvestri:2009hh,Clifton:2011jh,Tsujikawa:2013fta,2019ascl.soft10019T,Koyama:2015vza,Kase:2018aps}. The standard cosmological paradigm is the $\Lambda$CDM model, which consists of a cosmological constant $\Lambda$ as dark energy (DE) and cold dark matter (CDM).
Although the $\Lambda$CDM model is broadly consistent with a wide range of cosmological observations, including the cosmic microwave background (CMB) \cite{Planck:2018vyg}, supernovae type~Ia 
(SNe~Ia)\cite{Brout_2022,Abbott_2024}, baryon acoustic oscillations (BAO) \cite{BAOdr12,Adame_2025}, and large-scale structure data \cite{KiDS1000,DES2026}, it faces several persistent tensions as the precision of cosmological measurements increases \cite{Verde:2019ivm,Perivolaropoulos:2021jda,DiValentino:2020vvd,DiValentino:2021izs,Pantos:2026koc}. In particular, the $\Lambda$CDM prediction for the amplitude of matter clustering, characterized by the parameter $\sigma_8$, inferred from the Planck 2018 CMB data \cite{Planck:2018vyg}, is higher than that derived from low-redshift probes such as weak gravitational lensing, galaxy clustering or redshift-space distortions \cite{AbbottDESy3,KiDS1000,Li:2023tui,Wright:2025xka,DES:2026mkc,DES:2026fyc}. Although the level of tension can go from a few sigmas to being consistent, all probes seem to systematically give a value lower than the Planck 2018 result (see \cite{Pantos:2026koc} for a recent status report).  In addition, there exists a discrepancy in the inferred value of the present-day Hubble constant, $H_0$, between early- and late-Universe measurements. Although some (or all) of these tensions may ultimately be resolved through improved calibrations, the identification of previously unaccounted-for systematics, or the availability of additional data, they could also point to new physics beyond the standard cosmological model.

Among the various mechanisms proposed to address the $\sigma_8$ tension, elastic interactions (or pure momentum transfer) between CDM and a DE component have emerged as a compelling possibility. The underlying mechanism is that momentum transfer between the DM and a pressure-supported component, such as DE, generates an effective pressure for the coupled system that counteracts the gravitational collapse of the CDM structures, thereby reducing their clustering. If this interaction operates predominantly at late times, it can naturally reconcile the apparent deficit of CDM clustering at low redshift with that inferred from CMB observations. Such scenarios have been investigated in the context of elastic interactions between DE and DM \cite{Simpson:2010vh,Asghari:2019qld,Jimenez:2020ysu,Figueruelo:2021elm,Cardona:2022mdq,BeltranJimenez:2022irm,FiguerueloHernan:2023bjf,BeltranJimenez:2024lml,Jimenez:2024lmm,Poulin:2022sgp,Cruickshank:2025iig,Cruickshank:2025chm,BeltranJimenez:2025yad}, models involving scalar or vector fields \cite{Pourtsidou:2013nha,Boehmer:2015sha,Skordis:2015yra,Koivisto:2015qua,Pourtsidou:2016ico,Dutta:2017kch,Linton:2017ged,Kase:2019veo,Kase:2019mox,Chamings:2019kcl,Kase:2020hst,Amendola:2020ldb,DeFelice:2020icf,Linton:2021cgd,Liu:2023mwx,Aoki:2025bmj,Sahoo:2025cvz}, and interacting fluid models of DE and DM  \cite{BeltranJimenez:2020qdu,BeltranJimenez:2021wbq,Iosifidis:2024ksa,Jensko:2026taf}. We refer to \cite{Wang:2024vmw} for a review on interacting models.

A remarkable feature of models involving pure momentum transfer is that they do not modify the background cosmological evolution. Therefore, the potential alleviation of the $\sigma_8$ tension does not come at the expense of altering the background dynamics, which could otherwise worsen other tensions such as the Hubble tension. In fact, this property opens the possibility of simultaneously alleviating the Hubble tension by combining the pure momentum transfer with an additional mechanism capable of addressing the $H_0$ value. This consideration also motivates the study of non-elastic interactions involving energy transfer \cite{Kase:2019veo,Kase:2020hst,Amendola:2020ldb,Aoki:2025bmj}, 
since such interactions do affect the background cosmology. The resulting modification of the background evolution may help alleviating the Hubble tension without worsening the $\sigma_8$ tension, because the momentum transfer acts to disentangle the two tensions. Of course, this does not imply that such a disentanglement necessarily occurs in scenarios involving both energy and momentum transfer, but it nevertheless provides a promising framework.

A theoretically motivated realization of energy and momentum exchange in the dark sector is provided by a quintessence 
model of DE interacting with DM \cite{Amendola:2020ldb}. In this scenario, the momentum transfer induces an effective pressure for CDM, thereby suppressing the growth of structures, while the energy transfer gives rise to a $\phi$-matter-dominated epoch ($\phi$MDE) \cite{Amendola:1999er} that can help mitigate the $H_0$ tension. A recent analysis of a coupled quintessence model with energy and momentum exchange, using CMB data from Planck 2018 \cite{Aghanim:2019ame}, BAO measurements from SDSS DR12 \cite{BAOdr12}, Type Ia supernova data from Pantheon \cite{Brout_2022}, and DES data \cite{Abbott_2024}, reported evidence for a nonzero momentum-transfer coupling, suggesting a possible signature of dark-sector interactions even without including any $S_8$ or other low-redshift measurements \cite{Liu:2023mwx}. 
To better understand this unexpected result, we reanalyse the observational constraints on the coupled quintessence model using the latest data sets for the CMB, SNe Ia compilations, and BAO, also including low-redshift measurements of $S_8$. We find that evidence for a nonzero momentum-transfer coupling arises only when $S_8$ measurements are included in the statistical analysis. We further show that the results reported in \cite{Liu:2023mwx} were affected by a sampling error that led to misleading conclusions.

On the other hand, both momentum-exchange interactions and massive neutrinos share the property of suppressing the growth of structures at late times and on small scales. This raises the important question of whether their cosmological signatures are degenerate, and to what extent allowing for a varying neutrino mass could weaken or mask the detection of the dark-sector coupling. In this direction, the effects of massive neutrinos for a scenario with elastic interactions between CDM and DE was analysed in \cite{BeltranJimenez:2024lml} and it was found that the detection of a momentum transfer is robust against varying the neutrinos mass. Similarly, the recent study \cite{Pourtsidou2026} on a quintessence model with an exponential potential and featuring a pure momentum transfer also showed that the preference for a momentum transfer is largely insensitive to allowing for a varying neutrino mass.\footnote{The work \cite{Pourtsidou2026} updated the constraints from CMB, DESI-DR2 BAO and DES-Y5 SnIa on the quintessence model introduced in \cite{Pourtsidou:2013nha}  and it was found that the slope of the potential can be made compatible with the value conjectured from string theory in the presence of pure momentum transfer. Without low-redshift probes, the momentum transfer is compatible with the absence of interaction, but, interestingly, by fixing the slope of the potential to the lower bound predicted by string theory, a preference for a non-vanishing momentum transfer was found even without low-redshift data, which typically triggers the detection of momentum transfer.} These previous studies thus seem to indicate that a momentum transfer detection is not substantially compromised by varying the  neutrino mass.

The main goal of this work is to delve deeper into the potential degeneracy between massive neutrinos and momentum transfer to establish whether this a generic or a model-dependent feature. To that end, we will extend the previous analysis of the possible degeneracy between coupled quintessence with momentum transfer and massive neutrinos to the more general case including both energy and momentum exchange. The interacting DE-DM model considered in this work is the one proposed 
in \cite{Amendola:2020ldb}, with the effects of massive neutrinos included. We perform a Markov Chain Monte Carlo (MCMC) analysis using recent cosmological datasets, such as Planck CMB, SDSS, SNe~Ia, and $S_8$ measurements. In light of the latest BAO observations from DESI DR2 \cite{Adame_2025}, we also perform an extensive MCMC analysis combining these data with different SNe~Ia compilations, namely Pantheon+ \cite{Brout_2022} and 
DES~Y5 \cite{Abbott_2024}. Our main result will be that the evidence for a momentum transfer disappears when the mass of the neutrinos is allowed to vary, thus establishing the importance of considering a varying neutrino mass for these scenarios. Interestingly, the degradation of the momentum transfer detection is not exclusively due to the massive neutrinos, but it is a consequence of a new degeneracy between the neutrino mass and the DE-DM energy exchange and this explains why it was not observed in previous studies with pure momentum exchange.

This paper is organized as follows. In Sec.~\ref{backsec}, we revisit the interacting DE-DM model with both energy and momentum exchange and discuss the impact of the couplings and the neutrino mass on cosmological observables. In Sec.~\ref{MCMCresults}, we describe the methodology used to constrain the model with observational data and to place bounds on the couplings associated with energy and momentum transfers. In Sec.~\ref{results}, we summarize our results and discuss their interpretation.

\section{Interacting quintessence model}
\label{backsec}

%
\subsection{Theoretical framework}

The focus of this work will be a scenario where DE is described by a quintessence scalar field $\phi$ with an exponential potential, coupled to a CDM component such that energy and momentum are exchanged between the two dark components.
The total action of the model is given by
\be
{\cal S}=\int {\rm d}^4 x \sqrt{-g} \left[ \frac{\Mpl^2}{2}R+X-V_0  e^{-\lambda \phi/\Mpl}
-\left( e^{Q \phi/\Mpl}-1 \right)\rho_c+\beta Z^2 \right]
+{\cal S}_m\,,
\label{action}
\ee
where $g$ is the determinant of the metric tensor $g_{\mu \nu}$, $\Mpl$ is the reduced Planck mass, $R$ is the Ricci scalar, 
$X=-\partial_\mu\phi\partial^\mu\phi/2$ 
is the scalar kinetic term, 
and $V_0$ and $\lambda$ are constant parameters specifying the exponential potential. 
The fourth term in the action describes the energy transfer between the scalar field and the CDM component with energy density $\rho_c$, where the interaction strength is governed by the 
constant parameter $Q$.
The constant parameter $\beta$ measures the momentum transfer driven by the scalar product
\be
Z=\partial_{\mu} \phi\,u_c^{\mu}\,,
\ee
where $u_c^{\mu}$ is the 4-velocity of the CDM fluid. In the limit $Q \to 0$, the action (\ref{action}) recovers the momentum-transfer model proposed in \cite{Pourtsidou:2013nha}. 
As in the original paper, we focus 
on the interacting Lagrangian 
$\beta Z^2$, but it is possible to 
extend the momentum transfer to 
a more general form like 
$\beta X^{2-m}Z^m$ \cite{Kase:2019mox}, where 
$m$ is a constant.

The matter action ${\cal S}_m$ contains the contributions from CDM, baryons, and radiation. We model this matter sector as perfect fluids, which can be described by the Schutz-Sorkin action \cite{Schutz:1977df,Brown:1992kc,DeFelice:2009bx}. The equation-of-state (EOS) parameters for CDM, baryons, and radiation are denoted by $w_I$, with $I=c,b,r$, respectively. The radiation component has the usual relativistic EOS parameter $w_r=1/3$ and squared sound speed $c_r^2=1/3$. After CDM and baryons become nonrelativistic, their EOS parameters are approximately given by $w_c=0$ and $w_b=0$, with vanishing squared sound speeds $c_c^2=0$ and $c_b^2=0$.

The background and linear perturbation 
equations of motion on a spatially-flat 
Friedmann-Lema\^{i}tre-Robertson-Walker background were derived in Refs.~\cite{Amendola:2020ldb,Liu:2023mwx}. 
Hence, we refer the reader to these references for further details. 
The line element including the four scalar metric perturbations 
$\alpha$, $\chi$, $\zeta$, and $E$ is given by
\be
{\rm d}s^2=-(1+2\alpha) {\rm d}t^2
+2 \partial_i \chi {\rm d}t {\rm d}x^i
+a^2(t) \big[ (1+2\zeta) \delta_{ij}
+2\partial_i \partial_j E \big] {\rm d}x^i {\rm d}x^j\,,
\label{permet}
\ee
where $a(t)$ is the scale factor. We decompose the scalar field $\phi$ into 
a background part $\bar{\phi}(t)$ and a perturbation $\delta\phi(t,x^i)$ as
\be
\phi=\bar{\phi}(t)+\delta \phi (t,x^i)\,.
\ee
To alleviate the notation, we shall omit the bar in the following and $\phi$ will refer to the background value of the scalar field. Similarly, the energy densities of the fluids 
are decomposed as 
$\rho_I(t) + \delta\rho_I(t,x^i)$, 
where $\rho_I(t)$ is the background density 
and $\delta\rho_I$ denotes its perturbation.
The spatial components of the fluid four-velocities 
$u_{Ii}$ (with $i=1,2,3$) in perfect fluids 
are related to the scalar velocity potentials $v_I$ as 
$u_{Ii} = -\partial_i v_I$. 

To implement the background and perturbation 
equations in a Boltzmann solver, 
we introduce the conformal time 
$\tau=\int a^{-1}{\rm d}t$, 
and use a prime to denote derivatives with 
respect to $\tau$.
Then, the background equations of motion in the gravity sector 
are given by
\ba
& &
3\Mpl^2 \cH^2= \frac12 q_s \phi'^2
+a^2 \left( V_0 e^{-\lambda \phi/\Mpl}
+e^{Q\phi/\Mpl} \rho_c+\rho_b+\rho_r \right)\,,\label{EqBE01}\\
& &
2\Mpl^2 \left( \cH'-\cH^2 \right) 
= -q_s \phi'^2-a^2 \left( e^{Q\phi/\Mpl}\rho_c
+\rho_b+\frac{4}{3}\rho_r \right)\,,\label{EqBE02}
\ea
where 
\be
\cH=\frac{a'}{a}\,,
\qquad 
q_s=1+2\beta\,.
\ee
Here, the conformal Hubble parameter $\cH$ 
is related to the Hubble parameter 
$H=\dot{a}/a$ via $\cH=aH$, 
where a dot denotes derivatives with respect 
to $t$. 
The background equations for the scalar field and the perfect fluids are given, 
respectively, by 
\ba
& & 
q_s \left( \phi''+2 \cH \phi' \right)+\frac{a^2}{\Mpl} 
\left( Q\rho_c e^{Q\phi/\Mpl}
-\lambda V_0 e^{-\lambda \phi/\Mpl} 
\right)=0\,,
\label{EqBE03}\\
& & \rho_I'+3\cH \left( 1+w_I \right) 
\rho_I=0\,,
\label{EqBE04}
\ea
where $I=c,b,r$. 
We define the energy density $\rho_{\phi}$ and pressure $P_{\phi}$ of the scalar field, along with the 
coupled CDM density 
$\hat{\rho}_c$, as
\be
\rho_{\phi}=\frac{1}{2}\frac{q_s}{a^2} 
\phi'^2+V_0 e^{-\lambda \phi/\Mpl}\,,
\qquad 
P_{\phi}=\frac{1}{2}\frac{q_s}{a^2} 
\phi'^2-V_0 e^{-\lambda \phi/\Mpl}\,,
\qquad 
\hat{\rho}_c=e^{Q \phi/\Mpl}\rho_c\,.
\ee
Then, Eqs.~(\ref{EqBE03}) and (\ref{EqBE04}) can be rewritten as
\ba
& &
\rho_{\phi}'+3 \cH \left( \rho_{\phi}+P_{\phi} 
\right)=-\frac{Q \phi'}{\Mpl} 
\hat{\rho}_c\,, \\
& &
\hat{\rho}_c'+3\cH \hat{\rho}_c
=+\frac{Q \phi'}{\Mpl} \hat{\rho}_c\,.
\ea
These continuity equations indicate that the scalar field 
and CDM interact through an energy transfer characterized 
by the coupling constant $Q$. At the background level 
the coupling $\beta$ associated with the momentum transfer 
affects only the scalar-field kinetic term through 
$q_s = 1 + 2\beta$.

To study the evolution of perturbations, 
we choose the synchronous gauge 
\be
\alpha=0\,,\qquad \chi=0\,.
\ee
Instead of using $\zeta$ and $E$, 
we introduce the following variables: \cite{Ma:1995ey}
\be
\eta=-\zeta\,,\qquad 
h=-2k^2 E-6 \zeta\,.
\ee
We also define
\be
\theta_I=\frac{k^2}{a}v_I\,,\qquad 
\delta_I= \frac{\delta \rho_I}
{\rho_I}\,,
\ee
where $I=c,b,r$.
In Fourier space with comoving 
wavenumber ${\bm k}$, 
the linear perturbation equations take the following form:
\ba
\hspace{-0.3cm}
& &
k^2 \eta-\frac{\cH}{2} h'+\frac{a^2}{2\Mpl^2} 
\left[ \frac{q_s}{a^2}\phi' \delta \phi'
+\left( Q\rho_c e^{Q\phi/\Mpl}
-\lambda V_0 e^{-\lambda \phi/\Mpl} 
\right) \frac{\delta \phi}{\Mpl}
+e^{Q\phi/\Mpl} \rho_c \delta_c
+\rho_b \delta_b+\rho_r \delta_r \right]=0\,,\\
\hspace{-0.3cm}
& &
k^2 \eta'-\frac{a^2}{2\Mpl^2} 
\left[ \frac{k^2}{a^2}\phi' \delta \phi
+\left( \rho_c e^{Q\phi/\Mpl}+\frac{2\beta \phi'^2}{a^2} \right) 
\theta_c
+\rho_b \theta_b+\frac{4}{3}\rho_r \theta_r \right]=0\,,\\
\hspace{-0.3cm}
& &
\delta_c'+\theta_c+\frac{1}{2}h'=0\,,\\
\hspace{-0.3cm}
& &
\delta_b'+\theta_b+\frac{1}{2}h'=0\,,\\
\hspace{-0.3cm}
& &
\delta_r'+\frac{4}{3} \theta_r+\frac{2}{3}h'=0\,,\\
\hspace{-0.3cm}
& &
\theta_c'+\cH \theta_c-\frac{1}{q_s q_c \phi'^2 \Mpl^2} 
\biggl[ q_s (q_c-1)\phi' \Mpl k^2 \delta \phi' 
+\left\{ Q \phi'^2+a^2 (q_c-1)\lambda V_0 e^{-\lambda \phi/\Mpl}
\right\}k^2 \delta \phi \nonumber \\
\hspace{-0.3cm}
& &\qquad \qquad \qquad \qquad \qquad~
+\left\{ Q (q_s-2)\phi'^3+3q_s (q_c-1) \cH \phi'^2 \Mpl 
-2a^2(q_c-1)\phi' \lambda V_0 e^{-\lambda \phi/\Mpl} 
\right\}\theta_c \biggr]=0\,,
\label{EulerCDM}\\
\hspace{-0.3cm}
& &
\theta_b'+\cH \theta_b=0\,,\\
\hspace{-0.3cm}
& &
\theta_r'-\frac{k^2}{4}\delta_r=0\,,\\
\hspace{-0.3cm}
& &
\delta \phi''+2\cH \delta \phi'
+\frac{k^2 \Mpl^2+a^2 (\lambda^2 
V_0 e^{-\lambda \phi/\Mpl}+Q^2 \rho_c e^{Q\phi/\Mpl})}
{q_s \Mpl^2}\delta \phi
+\frac{\phi'}{2}h'+\frac{2\beta}{q_s}
\phi' \theta_c+\frac{a^2 Q\rho_c e^{Q\phi/\Mpl}}{q_s \Mpl}\delta_c
=0\,,\label{dphiddot}\\
\hspace{-0.3cm}
& &
h''+6 \eta''+2\cH (h'+6\eta')-2\eta k^2=0\,,
\ea
where $k=|{\bm k}|$, and
\be
q_c=1+\frac{4\beta x_1^2}
{\Omega_c}\,,
\ee
with 
\be
x_1=\frac{\phi'}{\sqrt{6}\Mpl \cH}\,,
\qquad 
\Omega_c=\frac{a^2 e^{Q \phi/\Mpl}\rho_c}
{3\Mpl^2 \cH^2}\,.
\ee
The effects of both energy and momentum transfer appear 
in the Euler equation~(\ref{EulerCDM}) 
as well as in the scalar-field perturbation equation~(\ref{dphiddot}).
To avoid ghost and Laplacian instabilities 
in dynamical perturbations, we require the 
following conditions
\be
q_s>0\,,\qquad q_c>0\,,\qquad 
c_s^2=\frac{1}{q_s} \left( 
1+\frac{8 \beta^2 x_1^2}
{4 \beta x_1^2+\Omega_c} \right)>0\,,
\label{stacon}
\ee
where $c_s^2$ is the squared sound speed 
of the scalar-field perturbation. 
As long as $\beta > 0$ and $\Omega_c > 0$, 
the three inequalities in Eq.~(\ref{stacon}) 
are automatically satisfied. This completes our brief review of the model and the relevant equations as well as the stability conditions that will serve as theoretical priors and we will proceed to analysing the evolution of the perturbations to obtain the corresponding constraints from data.

\subsection{Effects on cosmological observables}
\label{implementation}

Before confronting the model with observational information, 
we first review the impact of energy and momentum transfer on the evolution 
of cosmological perturbations to gain physical intuition on their effects. 
Furthermore, since we are 
interested in comparing the suppression of CDM clustering induced by the momentum transfer 
with that caused by massive neutrinos, we will also present a comparison between the two effects. 
The model has been implemented in the Boltzmann solver 
CAMB\footnote{\url{https://camb.info}. The modified version is available upon request.} 
(Code for Anisotropies in the Microwave Background) 
\cite{Lewis:1999bs,Howlett:2012mh}, using 
the perturbation equations 
in synchronous gauge, as presented in Sec.~\ref{backsec}. A pertinent cautionary remark to make at this point is that the interaction between DE and CDM prevents the possibility of completing the synchronous gauge by eliminating the CDM velocity perturbation. This is clearly seen in the CDM Euler equation \eqref{EulerCDM} where, even if we set $\theta_c=0$ at some initial time, the interactions mediated by $Q$ and $\beta$ will act as a source and, hence, a non-vanishing $\theta_c$ will be generated throughout the evolution. For the non-interacting case $Q=\beta=0$ (i.e. $q_s=q_c=1$), there is no source and $\theta_c$ remains vanishing. Thus, we will complement the usual adiabatic initial conditions with an adiabatic initial condition for $\theta_c$ as well. It is worth mentioning that the results do not depend on the choice of initial conditions for $\theta_c$ since, even if we set $\theta_c=0$ initially, the adiabatic solution for it is quickly reached.
For the varying neutrino mass case, we consider the coupled quintessence model with two massless 
neutrinos and one massive neutrino with mass $m_\nu$. The parameter $m_\nu$ will then be treated 
as a free parameter, together with $\beta$, $Q$, and $\lambda$, to study possible degeneracies.

\begin{figure}
\centering
\begin{tabular}{cc}
\includegraphics[width=0.49\textwidth]{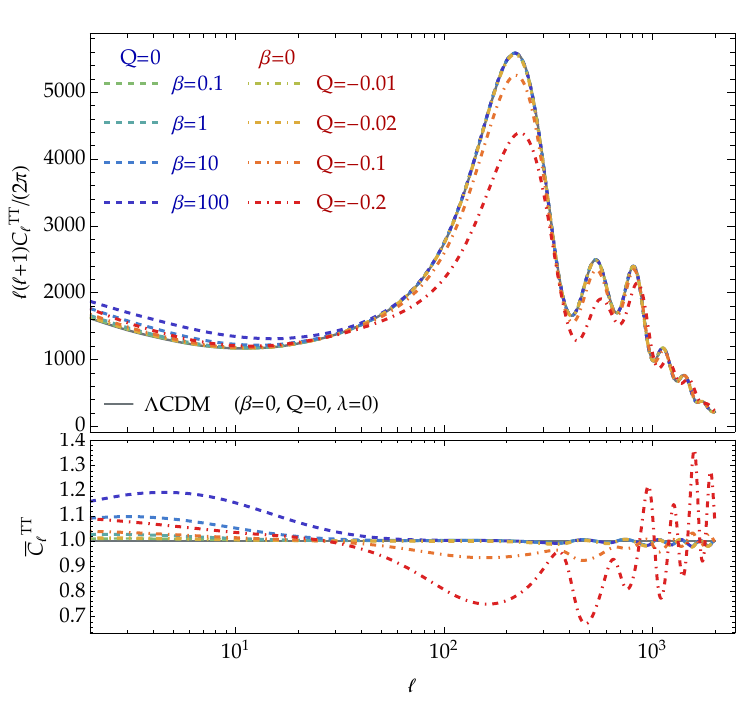}
\includegraphics[width=0.49\textwidth]{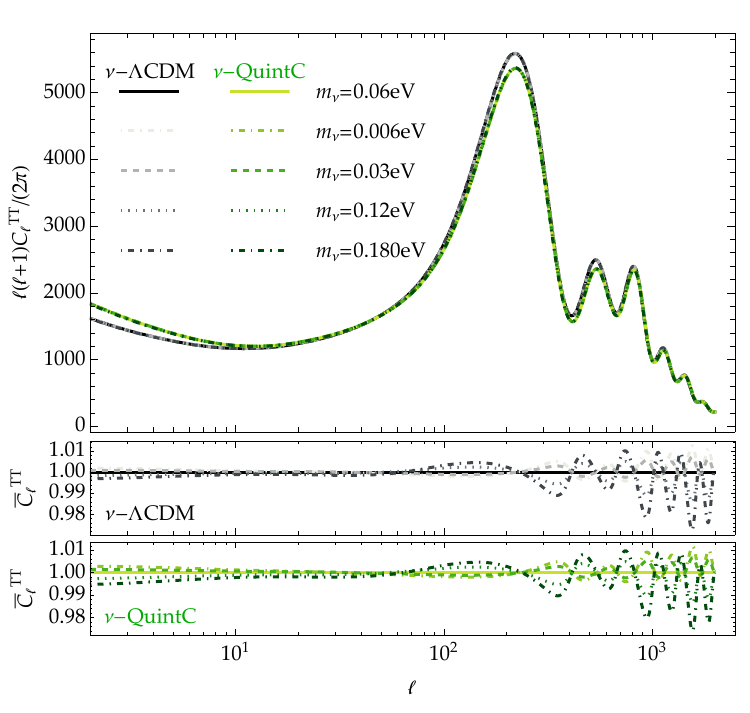}
\end{tabular}
\caption{Effects of coupled quintessence on the CMB angular power spectrum 
for temperature anisotropies. In the left panel, extreme cases with 
only momentum transfer ($Q=0$) and only energy transfer ($\beta=0$) 
are considered for fixed values of $\lambda=0.5$ and $m_\nu=0.06$ eV. 
In the right panel, we compare the impact of varying the neutrino 
mass $m_\nu$ in the $\nu$-QuintC ($\beta=1$, $Q=-0.1$, $\lambda=0.5$) 
and $\nu$-$\Lambda$CDM models. The relative ratios of 
$\bar{C}^{TT}_\ell$ are defined as 
$C^{TT}_\ell({\rm QuintC})/{C}^{TT}_\ell(\Lambda{\rm CDM})$ (left), 
${C}^{TT}_\ell(\nu\text{-}\Lambda{\rm CDM})/{C}^{TT}_\ell(\Lambda{\rm CDM})$ (right-top), and 
${C}^{TT}_\ell(\nu\text{-QuintC})/{C}^{TT}_\ell({\rm QuintC})$ 
(right-bottom).}
\label{plotsCMB}
\end{figure}

The effects of the DE-DM interaction 
on the CMB angular power spectrum 
of temperature anisotropies have been analysed 
in \cite{Liu:2023mwx}. 
Here we briefly review these effects for completeness and to compare them 
with the effects of massive neutrinos.
The corresponding results are shown 
in the left panel of Fig.~\ref{plotsCMB}. 
For vanishing energy transfer ($Q=0$), 
the angular power spectrum is 
mainly modified at large scales 
through the Integrated Sachs-Wolfe (ISW) 
effect. This is consistent with the fact 
that peculiar velocities become 
more relevant at late times, making 
the momentum transfer important at 
low redshift when DE also becomes dominant. 
When only energy transfer is present 
($\beta=0$), the coupling $Q$ leads 
to two main features relative to 
uncoupled quintessence: a shift of 
the acoustic peaks toward larger 
multipoles $\ell$ and a suppression 
of their amplitudes. 
This behaviour is consistent with the analysis 
presented in Refs.~\cite{Pettorino:2013oxa,Gomez-Valent:2020mqn}.
We therefore expect that 
$Q$ can be reasonably well constrained 
by CMB observations, whereas 
$\beta$ will be much less constrained 
by CMB data alone.

The effect of massive neutrinos on the CMB angular power spectrum in the coupled quintessence model (denoted as $\nu$-QuintC) is shown in the right panel of Fig.~\ref{plotsCMB}. In this figure, we fix the interaction parameters to $\beta=1$, $Q=-0.1$, and the slope of the potential to $\lambda =0.5$, while varying the neutrino mass $m_\nu$. As expected, the relative impact of massive neutrinos in the coupled quintessence model, with respect to the fiducial case 
$m_\nu=0.06\,{\rm eV}$, is similar to that in the $\nu$-$\Lambda$CDM model. The main effects arise through the late ISW effect ($\ell \lesssim 20$), the early ISW effect ($20 \lesssim \ell \lesssim 200$), and weak lensing ($\ell \gtrsim 200$) \cite{Workman:2022ynf}.

\begin{figure}
\centering
\begin{tabular}{cc}
\includegraphics[width=0.49\textwidth]{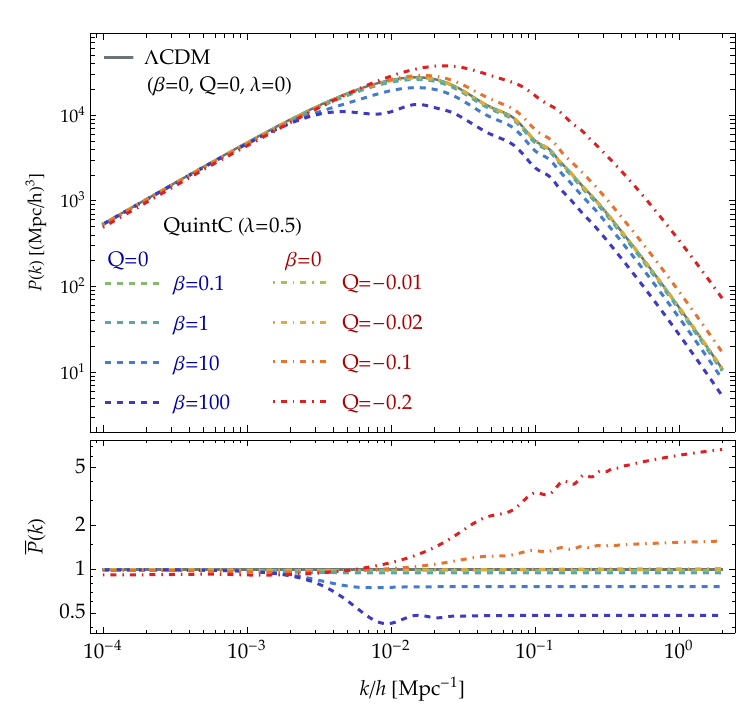}
\includegraphics[width=0.49\textwidth]{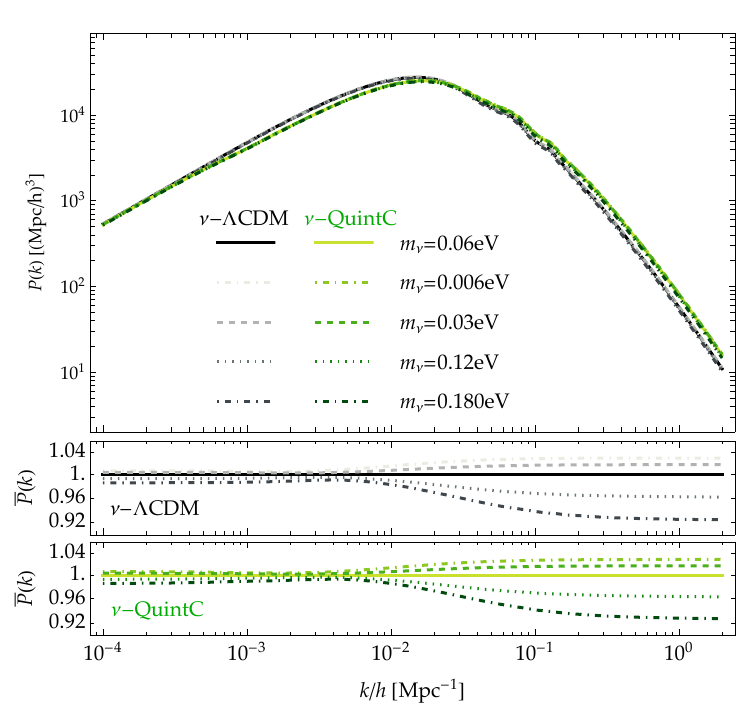}
\end{tabular}
\caption{Effects on the matter power spectrum in the coupled quintessence model and in the combined model including massive neutrinos. In the left panel, the extreme cases with only momentum transfer ($Q=0$) and only energy transfer ($\beta=0$) are considered for fixed $\lambda=0.5$ and $m_\nu=0.06\,{\rm eV}$. In the right panel, we compare the impact of varying the neutrino mass $m_\nu$ in the $\nu$-QuintC ($\beta=1$, $Q=-0.1$, $\lambda=0.5$) and $\nu$-$\Lambda$CDM models. In both panels, the relative differences are shown in the lower subpanels, where $\bar{P}(k)\equiv P(k)/P(k)_{\rm fiducial}$.}
\label{plotsPk}
\end{figure}

The effects of coupled quintessence on the matter power spectrum can be further analysed by considering the two extreme cases with only momentum transfer ($Q=0$) and only energy exchange ($\beta=0$), as shown in the left panel of Fig.~\ref{plotsPk}. In the former case, the momentum transfer leads to a suppression of power on small scales due to the slower growth of CDM density perturbations. This suppression originates from the effective pressure that the scalar field transfers to CDM through the momentum exchange.

On the other hand, the presence of energy exchange tends to enhance the growth of structures. This behaviour is consistent with the fact that the scalar field mediates an additional attractive interaction. The combined effect of both couplings depends on the relative strengths of the momentum and energy transfers. From the perspective of matter clustering, one may therefore expect a degeneracy between $Q$ and $\beta$, since they govern the energy and momentum exchange, respectively. However, this potential degeneracy is expected to be broken by CMB observations, which are significantly more sensitive to $Q$, as discussed above and can be clearly seen in Fig. \ref{plotsCMB}. From an observational standpoint, the momentum transfer is expected to play the dominant role in alleviating the $\sigma_8$ tension. Although one might anticipate that an energy exchange could help address the Hubble tension, we will show later that this is not the case in the present model. Hence, the suppression of structures driven by the momentum transfer remains the dominant effect.

As explained above, the most relevant feature for this work is the suppression of the matter power spectrum on small scales, since this is what allows the $\sigma_8$ tension to be alleviated. In fact, measurements of $\sigma_8$ (or $S_8$) are those most likely to signal a detection of a pure momentum transfer (see, e.g., \cite{BeltranJimenez:2021wbq} for a detailed discussion). Since massive neutrinos also lead to a suppression of the matter power spectrum on small scales, a pertinent question is whether varying the neutrino mass could affect the significance of a momentum-transfer detection. This issue has already been investigated in another scenario involving momentum transfer in \cite{BeltranJimenez:2024lml}, where it was shown that the detection of the momentum transfer is robust against variations in the neutrino mass. In this work, we perform an analogous analysis for the coupled quintessence model to assess whether such robustness is a generic feature or depends on the specific realization of the momentum-transfer scenario. In the right panel of Fig.~\ref{plotsPk}, we show the effect of varying the neutrino mass on the matter power spectrum, where the aforementioned suppression on small scales can be clearly observed. This suppression could potentially degrade a detection of $\beta$. Although the scale dependence of the two effects differs, measurements of $\sigma_8$, being an integrated quantity, are not sensitive to the detailed scale dependence and could therefore lead to a degeneracy between $m_\nu$ and $\beta$. Determining whether such a degeneracy is present constitutes one of the main goals of this work. 
Let us note, however, that this degeneracy may be broken by the effects of massive neutrinos on the background evolution and the CMB, since $\beta$ barely affects these observables.

After this brief review of the model and its effects on the CMB and the matter power spectrum, we now proceed to the confrontation with data.

\section{Confrontation with data}
\label{MCMCresults}

\subsection{Models and datasets}
\label{Sec:Modelanddata}

In this section, we confront the coupled quintessence model described in the preceding sections with observational data by performing MCMC analyses. We begin by revisiting previously reported results and by investigating the prominent role of low-redshift measurements in obtaining a significant detection of the pure momentum-transfer interaction parameter. 
We then allow the neutrino mass to vary to study its impact on the results and the possible degeneracies that may arise, which is the main focus of this work.

We will consider the following cosmological scenarios: 
\begin{enumerate}[label={(\alph*)}]
\item {\bf QuintC}: The interacting quintessence model described in Sec.~\ref{backsec}. In addition to the standard cosmological parameters---namely the present-day baryon density $\Omega_{b0}h^2$, the present-day CDM density $\Omega_{c0}h^2$, the angular size of the comoving sound horizon at recombination $100\,\theta_s$, the amplitude of primordial scalar perturbations $\log(10^{10}A_s)$, the scalar spectral index $n_s$, and the optical depth to reionization $\tau_{\rm reio}$---this model introduces three additional parameters to be sampled: the coupling constants $Q$ and $\beta$, which describe the energy and momentum interactions, respectively, and $\lambda$, which characterizes the slope of the exponential potential.

\item {\bf $\nu$-QuintC}: The interacting quintessence model QuintC with two massless neutrinos and one massive neutrino of mass $m_\nu$. In addition to $Q$, $\beta$, and $\lambda$, the neutrino mass $m_\nu$ is treated as an additional free parameter of the model.
\end{enumerate}
For comparison, we shall also consider the standard $\Lambda$CDM model and its extension with varying neutrinos mass $\nu-\Lambda$CDM. For the new parameters of the QuintC model and the neutrino mass we will use flat priors in the following ranges:
\be
{\log_{{10}}\,\beta } \in [-3,3],
\quad \lambda \in [0,1.5],\quad  
Q \in [-0.08,0],\quad
\text{and}\quad m_\nu \in 
[0,5]~{\rm eV}\,.
\ee
Importantly, we adopt a flat prior on the logarithm of the momentum-transfer parameter $\beta$. This choice becomes relevant when including $S_8$ measurements in the analysis aimed at detecting a non-vanishing $\beta$, and in fact allows us to correct an improper detection reported in \cite{Liu:2023mwx}. Furthermore, this choice automatically takes into account the theoretical prior $\beta\geq0$. As we will show, our results are not prior-dominated, indicating that this choice of prior is reasonable. The slope parameter $\lambda$ is chosen to be positive without loss of generality. Since we are interested in the case where the $\phi$MDE \cite{Amendola:1999er} is present, we restrict the coupling $Q$ to the range $Q \le 0$ in the MCMC analysis, following Ref.~\cite{Liu:2023mwx}.

In the following, we describe the different combinations of datasets used in our MCMC analyses. These choices are motivated by the need to facilitate comparison with previous results and to assess the impact of different datasets on our constraints. We therefore consider the following three baseline datasets:
\begin{itemize}
\item {\bf Baseline I}: Planck 2018 [TTTEEE] \cite{Aghanim:2019ame}; Pantheon+ SNe Ia compilation \cite{Brout_2022}; and BAO from SDSS DR12 \cite{BAOdr12}.
    
\item {\bf Baseline II}: Planck 2018 [TTTEEE] \cite{Aghanim:2019ame}; Pantheon+ SNe Ia compilation \cite{Brout_2022}; and  BAO from DESI DR2 \cite{Adame_2025}.

\item {\bf Baseline III}: Planck 2018 [TTTEEE] \cite{Aghanim:2019ame}; 
DES-Y5 SNe~Ia \cite{Abbott_2024}; and BAO from DESI DR2 \cite{Adame_2025}.

\end{itemize}
These three baseline datasets will in turn allow us to study the impact of using different BAO and supernova datasets. This is particularly relevant in light of the latest DESI BAO results, which have pointed to a possible dynamical DE component (although this claim still requires careful assessment), as well as the observed dependence of cosmological parameter constraints on the choice of supernova dataset. We restrict our analysis to Planck CMB data and do not include ACT or SPT measurements. Our goal is not to obtain the most precise and comprehensive constraints on the model, but rather to investigate the effect of massive neutrinos. As a secondary goal, we also aim to compare our results with those reported in \cite{Liu:2023mwx}, and therefore we adopt the same CMB dataset.

We will also consider extensions of the above Baseline datasets by supplementing them with a measurement of $S_{8,\rm obs}$, added as a Gaussian likelihood of the following form:
\begin{equation}
\log {\cal L}_{S_8} =
-\frac{\left(S_{8,{\rm th}}-S_{8,{\rm obs}}\right)^2}{2\sigma_{S_8}^2}\,,
\end{equation}
where $S_{8,{\rm th}}$ is the theoretical prediction and $\sigma_{S_8}$ is the uncertainty of the considered measurement. We use the following two independent measurements of $S_{8,{\rm obs}}$ :
\begin{itemize}
\item Planck SZ 2013:
$S_{8,{\rm SZ}} \equiv \sigma_8 (\Omega_{m0}/0.27)^{0.3}
= 0.782 \pm 0.010$ \cite{refId0}.
\item DES-Y3:
$S_{8,{\rm DES}} \equiv \sigma_8 (\Omega_{m0}/0.3)^{0.5}
= 0.776 \pm 0.017$ 
\cite{PhysRevD.107.023531},
\end{itemize}
where $\Omega_{m0}=
\Omega_{c0}+\Omega_{b0}$.
We consider these two values to highlight the importance of this measurement for the detection of a nonvanishing $\beta$. Using these measurements as a Gaussian likelihood involves some caveats, which have already been discussed in, e.g., \cite{BeltranJimenez:2021wbq}. In this work, we assume that this method provides a reasonably good approximation for our purposes and adopt it as the commonly employed approach, although a more rigorous assessment of its general validity would be desirable.
 
In order to obtain the observational constraints, we have implemented a modified version of CAMB into COBAYA\footnote{\url{https://cobaya.readthedocs.io/en/latest/\#}} \cite{Torrado_2021,2019ascl.soft10019T}. We have also explored an implementation in COSMOMC\footnote{\url{http://cosmologist.info/cosmomc}} \cite{Lewis:2002ah,Lewis:2013hha}, which we used to run several tests, finding good agreement between the results obtained with both codes. In all cases, we follow the Gelman-Rubin criterion \cite{GelmanRubin}, $|R-1| < 0.01$, to ensure the convergence of the chains. The statistical analysis and contour plots were produced using the GetDist\footnote{\url{https://getdist.readthedocs.io/en/latest/index.html}} package \cite{Lewis:2019xzd}.

\subsection{Constraints on the 
Quint-C model and the role of $S_8$}

\begin{figure}
\centering
\includegraphics[width=1.\textwidth]{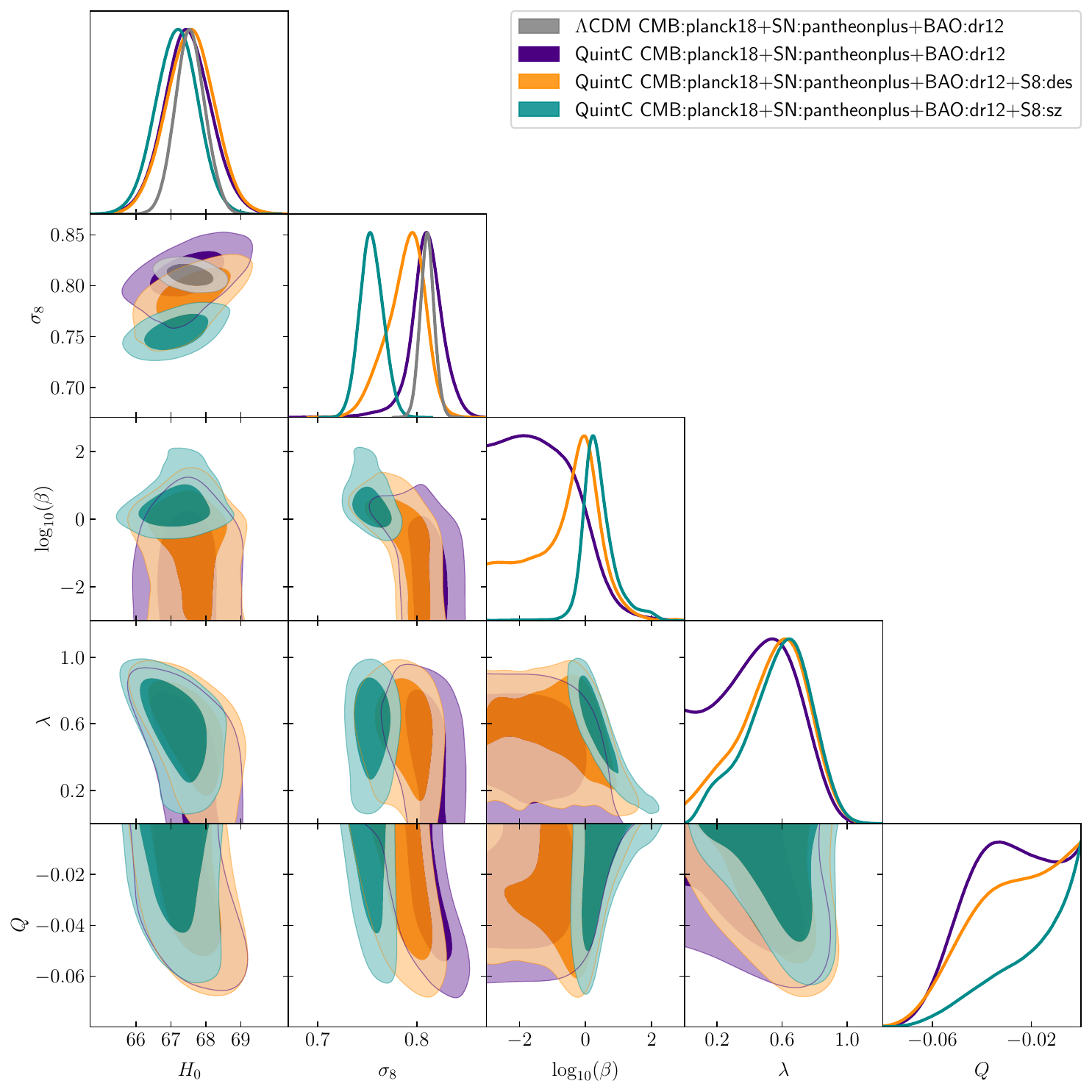}
\caption{Constraints on the QuintC model, together with the $\Lambda$CDM model, derived from the combination 
of several datasets. 
In the two-dimensional planes, the inner and outer contours correspond to the $1\sigma$ and $2\sigma$ confidence 
regions, respectively.
}
\label{QuintCtriangleS8}
\end{figure}

We start by comparing our results with previous results in the literature, in particular with the analysis performed in \cite{Liu:2023mwx}, 
as well as with those of \cite{Pourtsidou:2016ico,Pourtsidou2026} for scenarios without energy exchange. 
The MCMC results are presented in Table~\ref{tableMCMCbase}, and the corresponding triangle plots showing the confidence regions are displayed in Fig.~\ref{QuintCtriangleS8}. 

\begin{table}
\begin{center}
\renewcommand{\arraystretch}{1.8}
\begin{tabular}{ |c|c|c|c|c| } 
\hline
\hline
\centering
	 &QuintC Baseline I & QuintC Baseline I+$S_{8,\text{SZ}}$ & $\nu$-QuintC Baseline I & $\nu$-QuintC Baseline I+$S_{8,\text{SZ}}$ \\ 
     \hline
	Parameter & mean\,$\pm\sigma\pm 2\sigma$ & mean\,$\pm\sigma\pm 2\sigma$  &  mean\,$\pm\sigma \pm 2\sigma$ & mean\,$\pm\sigma \pm 2\sigma$\\
	\hline \hline 
$100\Omega_{b0} h^2$ & $2.24_{-0.01}^{+0.01} {}_{-0.03}^{+0.03}$ & $2.24_{-0.01}^{+0.01}{} _{-0.03}^{+0.03}$ & $2.24_{-0.01}^{+0.01} {}_{-0.03}^{+0.03}$ & $2.24_{-0.02}^{+0.02} {}_{-0.03}^{+0.03}$ \\\hline
$\Omega_{c0}h^2$  & $0.118_{-0.001}^{+0.002} {}_{-0.003}^{+0.003}$ & $0.118_{-0.001}^{+0.002}  {}_{-0.003}^{+0.003}$ & $0.118_{-0.001}^{+0.002} {}_{-0.004}^{+0.003}$  & $0.117_{-0.001}^{+0.003} {}_{-0.005}^{+0.004}$ \\\hline 
$n_{s}$ & $0.966_{-0.004}^{+0.004} {}_{-0.008}^{+0.008}$ & $0.969_{-0.004}^{+0.004} {}_{-0.008}^{+0.008}$ & $0.966_{-0.004}^{+0.004} {}_{-0.008}^{+0.008}$ & $0.968_{-0.004}^{+0.004} {}_{-0.008}^{+0.008}$ \\ \hline  
$\log(10^{10}A_{s})$ & $3.05_{-0.02}^{+0.02} {}_{-0.03}^{+0.03}$ & $3.04_{-0.02}^{+0.02} {}_{-0.03}^{+0.03}$  & ${3.05}_{-0.02}^{+0.02} {}_{-0.03}^{+0.03}$  & $3.04_{-0.02}^{+0.02} {}_{-0.03}^{+0.03}$ \\ \hline  
$\tau_{\rm reio}$  & $0.056_{-0.008}^{+0.007} {}_{-0.015}^{+0.016}$ & $0.054_{-0.008}^{+0.008} {}_{-0.015}^{+0.016}$ &  $0.056_{-0.008}^{+0.007} {}_{-0.015}^{+0.016}$ & $0.054_{-0.008}^{+0.008} {}_{-0.015}^{+0.016}$  \\ \hline  
$100~\theta_{s}$  & $1.0410_{-0.0003}^{+0.0003} {}_{-0.0006}^{+0.0006}$ & $1.0411_{-0.0003}^{+0.0003} {}_{-0.0006}^{+0.0006}$ & $1.0411_{-0.0003}^{+0.0003} {}_{-0.0006}^{+0.0006}$ & $1.0412_{-0.0003}^{+0.0003} {}_{-0.0006}^{+0.0006}$ \\ \hline \hline
$H_0$ [km/s/Mpc] & $67.48_{-0.66}^{+0.66} {}_{-1.34}^{+1.4}$  & $67.17_{-0.61}^{+0.6} {} _{-1.22}^{+1.16}$ & $67.5_{-0.7}^{+0.68} {}_{-1.41}^{+1.42}$   & $67.2_{-0.66}^{+0.66} {}_{-1.31}^{+1.28}$ \\ \hline 
$\sigma_8$  & $0.809_{-0.014}^{+0.016} {}_{-0.034}^{+0.037}$  & $0.754_{-0.012}^{+0.011} {}_{-0.022}^{+0.024}$ & $0.81_{-0.014}^{+0.02} {}_{-0.039}^{+0.036}$   & $0.756_{-0.012}^{+0.012} {}_{-0.023}^{+0.024}$ \\ \hline 
$\Omega_{m0}$  & $0.31_{-0.008}^{+0.009} {}_{-0.017}^{+0.015}$ & $0.312_{-0.007}^{+0.008} {}_{-0.016}^{+0.015}$ & $0.309_{-0.008}^{+0.009} {}_{-0.017}^{+0.017}$ & $0.311_{-0.008}^{+0.008} {}_{-0.017}^{+0.016}$ \\ \hline  
\hline
${\log_{{10}}\,\beta }$ & ${-1.32}^{+0.54} {}_{}^{+1.71}$  & $0.41_{-0.53}^{+0.30} {}_{-0.81}^{+1.29}$ & $-1.35_{}^{+0.53} {}_{}^{+1.66}$   & $-0.01_{-0.38}^{+0.95} {}_{}^{+1.27}$\\ \hline 
${\lambda}$ & $0.44_{-0.23}^{+0.28} {}_{}^{+0.37}$ & $0.57_{-0.16}^{+0.24} {}_{-0.43}^{+0.34}$ & $0.46_{-0.22}^{+0.29} {}_{}^{+0.36448}$   & $0.54_{-0.17}^{+0.26} {}_{-0.45}^{+0.37}$ \\ \hline 
${Q}$  & $-0.028_{-0.009}^{} {}_{-0.028}^{\ }$ & $-0.02_{-0.006}^{\ } {}_{-0.031}^{\ }$ & $-0.029_{-0.009}^{\ } {}_{-0.031}^{\ }$   & $-0.029_{-0.010}^{} {} _{-0.451}^{}$  \\ \hline
$m_{\nu}$  & $-$ & $-$ & $0.06_{}^{+0.009} {}_{}^{+0.118}$ & $0.126_{}^{+0.012} {}_{}^{+0.291}$  \\ \hline
\hline
\end{tabular}
\end{center}
\caption{MCMC results for the mean values and the $1\sigma$ ($68\,\%$ CL) and $2\sigma$ ($95\,\%$ CL) upper and lower limits of the cosmological and derived parameters for the QuintC and 
$\nu$-QuintC models using the Baseline I and Baseline I+$S_8$ datasets (with the Planck SZ 2013 measurement of $S_8$). The priors for the additional model parameters are ${\log_{10}\beta} \in [-3,3]$, $\lambda \in [0,1.5]$, and $Q \in [-0.08,0]$, together with 
$m_\nu \in [0,5]$~eV for 
the $\nu$-QuintC model. 
}
\label{tableMCMCbase}
\end{table}

The first important result is that the momentum transfer parameter $\beta$ is compatible with 0 (i.e., no momentum transfer) at $1\sigma$ when using the Baseline I data (purple regions). However, the inclusion of the $S_{8,\text{SZ}}$ measurement from Planck SZ (Baseline I+$S_{8,\text{SZ}}$, green regions) tends to favor a nonvanishing momentum transfer, with $\log_{10}\beta = 0.41^{+0.30}_{-0.53}$, and the noninteracting case being excluded at more than $2\sigma$. The pure momentum-transfer scenario with $Q=0$ was analysed in \cite{Pourtsidou:2016ico}, and our findings are compatible with their results. Furthermore, we find that the inclusion of energy transfer mediated by $Q$ does not spoil the detection of $\beta$. Indeed, as can be seen in the $Q$-$\log_{10}\beta$ plane, the presence of energy exchange does not significantly affect the constraints on $\beta$. This corroborates our expectation that the two interaction parameters do not lead to a significant degeneracy.

\begin{figure}
\centering
\includegraphics[width=1.\textwidth]{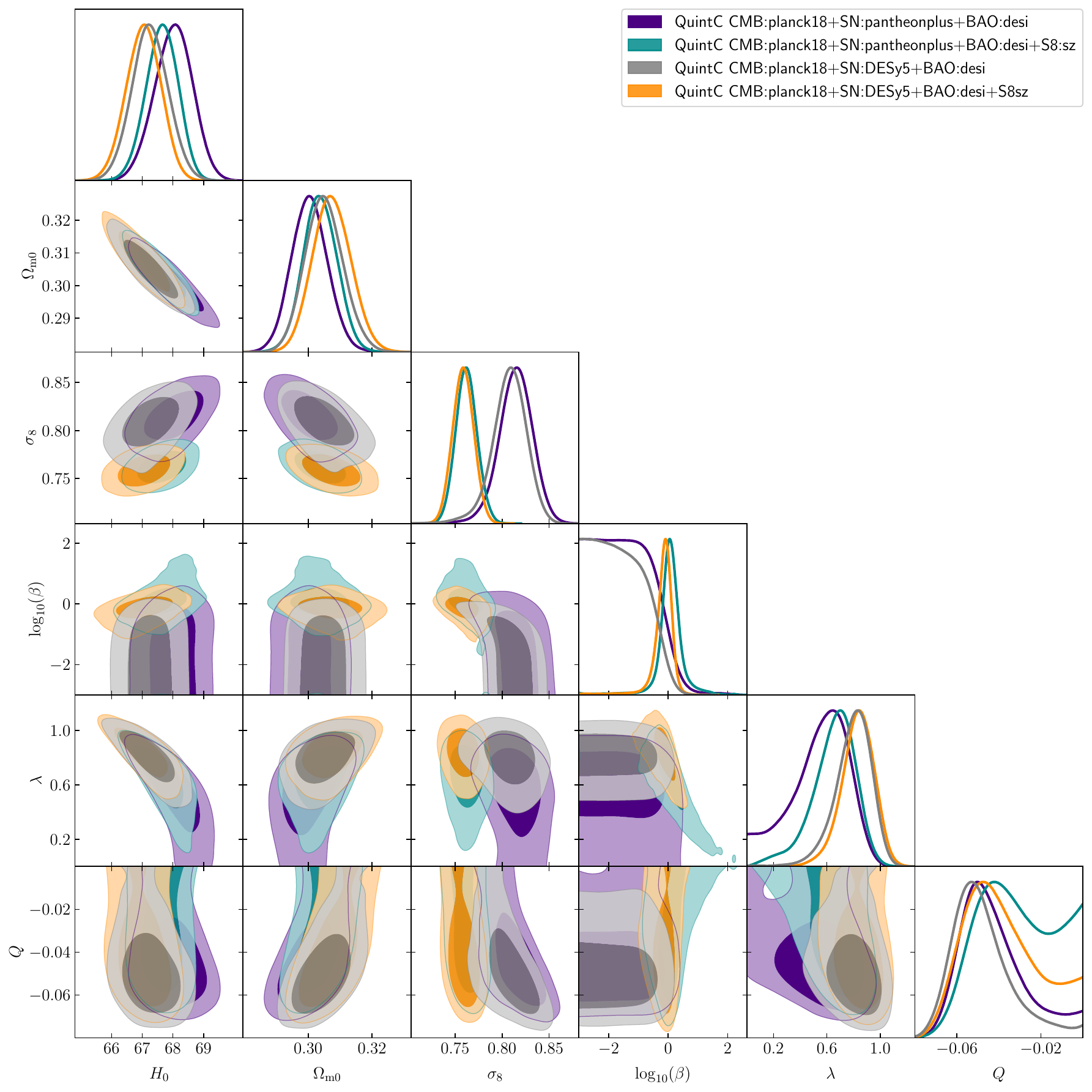}
\caption{Constraints on the QuintC model obtained from the combination of several datasets, including BAO measurements from DESI-DR2 \cite{Adame_2025} and SNe~Ia from DES-Y5 \cite{Abbott_2024}.
}
\label{QuintCtriangleDESIall}
\end{figure}
\begin{table}
\begin{center}
\renewcommand{\arraystretch}{1.8}%
\begin{tabular}{ |c|c|c|c|c| } 
	\hline
	\hline
	\centering
	QuintC &  Baseline II & Baseline II+$S_{8}$ &Baseline III &Baseline III+$S_{8}$ \\ \hline
	Parameter & mean\,$\pm\sigma\pm 2\sigma$ & mean\,$\pm\sigma \pm\sigma$ & mean\,$\pm\sigma \pm 2\sigma$ & mean\,$\pm\sigma \pm 2\sigma$\\
	\hline \hline 
$100\Omega_{b0} h^2$  & 
$2.24_{-0.02}^{+0.02} {}_{-0.03}^{+0.03}$  &
$2.25_{-0.01}^{+0.02} {}_{-0.03}^{+0.03}$ & 
$2.24_{-0.02}^{+0.02} {}_{-0.03}^{+0.03}$ 
& $2.24_{-0.02}^{+0.02} {}_{-0.04}^{+0.03}$  
\\ \hline
$\Omega_{c0}h^2$  &
$0.116_{-0.001}^{+0.001} {}_{-0.002}^{+0.002} $ & $0.116_{-0.001}^{+0.001} {}_{-0.002}^{+0.002}$ & 
$0.115_{-0.001}^{+0.001} {}_{-0.003}^{+0.003}$ & 
$0.115_{-0.001}^{+0.001} {}_{-0.003}^{+0.002}$
\\ \hline 
$n_{s}$  
& $0.968_{-0.004}^{+0.004} {}_{-0.008}^{+0.008}$  & $0.971_{-0.003}^{+0.003} {}_{-0.007}^{+0.007}$& 
$0.968_{-0.004}^{+0.004} {}_{-0.008}^{+0.008}$& 
$0.971_{-0.004}^{+0.004} {}_{-0.007}^{+0.007}$
\\ \hline  
$\log(10^{10}A_{s})$ & 
$3.05_{-0.02}^{+0.02} {}_{-0.032}^{+0.034}$ & 
$3.04_{-0.016}^{+0.016} {}_{-0.033}^{+0.033}$ & 
$3.05_{-0.02}^{+0.02} {}_{-0.031}^{+0.033}$ & 
$3.04_{-0.02}^{+0.02} {}_{-0.033}^{+0.033}$ 
\\ \hline  
$\tau_{\rm reio}$  &  
$0.056_{-0.008}^{+0.008} {}_{-0.016}^{+0.017}$ & $0.055_{-0.008}^{+0.008} {}_{-0.016}^{+0.016}$&
$0.056_{-0.008}^{+0.008} {}_{-0.016}^{+0.017}$ & 
$0.054_{-0.008}^{+0.008} {}_{-0.016}^{+0.016}$ 
\\ \hline  
$100~\theta_{s}$  &
${1.0412}_{-0.0003}^{+0.0003} {}_{-0.0006}^{+0.0006}$ & $1.0413_{-0.0003}^{+0.0003} {}_{-0.0006}^{+0.0006}$ & $1.0412_{-0.0003}^{+0.0003} {}_{-0.0006}^{+0.0006}$ & $1.0413_{-0.0003}^{+0.0003} {}_{-0.0006}^{+0.0006}$
\\ \hline  \hline
$H_0$ [km/s/Mpc] & 
$68.01_{-0.62}^{+0.62} {}_{-1.23}^{+1.19}$ & 
$67.63_{-0.5}^{+0.56} {}_{-1.09}^{+0.99}$ & 
$67.26_{-0.59}^{+0.59} {}_{-1.16}^{+1.18}$& 
$67.04_{-0.56}^{+0.57} {}_{-1.12}^{+1.07}$
\\ \hline  
$\sigma_8$ 
& $0.814_{-0.017}^{+0.018} {}_{-0.036}^{+0.036}$  & $0.762_{-0.011}^{-0.011} {}_{-0.022}^{+0.022}$&
$0.807_{-0.015}^{+0.02} {}_{-0.038}^{+0.037}$ &
$0.759_{-0.011}^{+0.011}{}_{-0.021}^{+0.022} $
\\ \hline  
$\Omega_{m0}$ & 
$0.300_{-0.006}^{+0.006} {}_{-0.011}^{+0.012}$ & $0.304_{-0.005}^{+0.005} {}_{-0.010}^{+0.011}$ & $0.305_{-0.006}^{+0.006} {}_{-0.011}^{+0.012}$& 
$0.308_{-0.006}^{+0.006} {}_{-0.011}^{+0.012}$ 
\\ \hline\hline
${\log_{{10}}\,\beta }$ & 
${-1.48}_{}^{+0.5} {}_{}^{+1.47}$ & 
$0.07_{-0.29}^{+0.3} {}_{-0.94}^{+0.98} $ & 
$-1.64_{}^{+0.46} {}_{}^{+1.4}$ & 
$-0.19_{-0.16}^{+0.33} {}_{-0.74}^{+0.67}$ 
\\ \hline
${\lambda}$ & 
$0.549_{-0.143}^{+0.260} {}_{-0.484}^{+0.34}$ &
$0.648_{-0.118}^{+0.19} {}_{-0.36}^{+0.31}$ & 
$0.79_{-0.102}^{+0.159} {}_{-0.29}^{+0.26}$& 
$0.827_{-0.103}^{+0.138} {}_{-0.25}^{+0.24}$\\ \hline
${Q}$    & 
$-0.043_{-0.018}^{+0.009} {}_{-0.021}^{ }$ & 
$-0.032_{-0.011}^{} {}_{-0.027}^{ }$ & 
$-0.047_{-0.017}^{+0.009} {}_{-0.02}^{ }$& $-0.038_{-0.024}^{+0.013} {}_{-0.026}^{}$ \\ \hline
\hline 
\end{tabular} 
\end{center}
\caption{MCMC results for the mean values and the $1\sigma$ (68\,\% CL) 
and $2\sigma$ (95\,\% CL) upper and lower limits of the cosmological and derived parameters in the QuintC model, obtained using the Baseline II and Baseline III datasets, as well as the combinations Baseline II+$S_8$ and Baseline III+$S_8$, where the SZ measurement of $S_8$ is included. Flat priors are assumed for the additional model parameters in the ranges ${\log_{10}\beta} \in [-3,3]$, $\lambda \in [0,1.5]$, and $Q \in [-0.08,0]$.
}
\label{tableMCMCdesi}
\end{table}

On the other hand, our results do not agree with the previous analysis performed in \cite{Liu:2023mwx} for the same coupled quintessence model. In that work, 
a detection of the momentum-transfer parameter, $\ln \beta = -1.1546^{+0.4925}_{-0.0639}$ ($68\%$ CL), was reported using only the Baseline~I data, without including low-redshift probes such as measurements of $S_8$ or other observables (e.g., weak lensing) that could potentially drive such a detection. As discussed above, the Baseline I dataset can only place an upper bound on the momentum transfer and does not lead to a detection. We have traced this discrepancy to an improper sampling of the momentum-transfer parameter in \cite{Liu:2023mwx}. Once the sampling is performed correctly, we find full agreement with the results obtained in this work. 

As discussed above, the inclusion of $S_{8,\text{SZ}}$ is crucial for the detection of $\beta$, so it is natural to ask how robust this detection is with respect to the specific choice of the $S_8$ measurement. In previous works with similar scenarios, it was found that the detection is robust against replacing the $S_8$ measurement with, e.g., $S_{8,\text{DES}}$ (see, for instance, Fig.~4 in \cite{Jimenez:2024lmm}). Although we routinely test several of these measurements in our works, we usually report only one because the results do not depend strongly on the particular choice of $S_8$. However, the coupled quintessence model analysed in this work does not follow this trend. In fact, we find that the detection is quite sensitive to the adopted $S_8$ measurement. For this reason, we also report the results corresponding to Baseline I+$S_{8,\text{DES}}$ (orange regions) in Fig.~\ref{QuintCtriangleS8}. By comparing the orange (Baseline I+$S_{8,\text{DES}}$) and green (Baseline I+$S_{8,\text{SZ}}$) confidence regions in Fig.~\ref{QuintCtriangleS8}, we see that a detection of $\beta$ occurs only in the latter case. This difference can be interpreted in terms of the relative errors of the two measurements, with the DES-Y3 relative uncertainty being approximately a factor of $\sim 2$ larger than that of Planck SZ. While this difference appears to be unimportant in other scenarios \cite{BeltranJimenez:2021wbq}, it seems to play a crucial role for the coupled quintessence model considered here.

In the latest DES-Y6 data release, the reported value of $S_8$ (from the $3\times2$ pt probe) has a slightly smaller uncertainty, although the mean value is somewhat larger. It is therefore unclear whether the detection would be recovered using this updated measurement. In particular, although some tension with Planck 2018 remains, it is reduced relative to the DES-Y3 data release. The DES-Y6 weak-lensing value of $S_8$ \cite{DES:2026mkc} is very similar to the DES-Y3 value adopted here, so we do not expect significant changes.

In any case, we emphasize that the goal of this paper is not to perform a comprehensive analysis of the constraints from different surveys on the coupled quintessence model, but rather to investigate whether a potential detection of $\beta$ can be spoiled by allowing for a varying neutrino mass. Before addressing this question, however, it is interesting to examine the impact of adopting different BAO and SNe~Ia datasets.

\subsection{DESI data and 
a detection of energy transfer 
for the QuintC model}
\label{subsecDESI}

After revisiting (and correcting) previous results in the literature, we explore how the constraints on the model parameters are affected when using different BAO and supernova datasets. We first study this effect using BAO data from DESI-DR2 while maintaining Planck 2018 and Pantheon+ (Baseline II). The results are shown in Fig.~\ref{QuintCtriangleDESIall} and detailed in the third and fourth columns of Table~\ref{tableMCMCdesi}.

Interestingly, using DESI-DR2 BAO data leads to a mild evidence for a nonvanishing value of the parameter $Q$ (purple and green regions). This preference becomes slightly stronger when the supernova dataset is also changed to DES-Y5 (gray and yellow regions), i.e., when using the Planck 2018+DESI-DR2+DES-Y5 combination (Baseline III). Of course, the statistical significance of this preference remains weak: the value $Q=0$ lies outside the $1\sigma$ region but well within the $2\sigma$ region. The feature we want to highlight is not this very mild evidence itfself, but the fact that the SDSS DR12 BAO data is instead perfectly compatible with $Q=0$ even at $1\sigma$ as can be seen in Fig. \ref{QuintCtriangleS8}. This result become even more intriguing when considering the parameter $\lambda$. In particular, the Baseline III combination increases the preference for a nonvanishing value of $\lambda$ compared to the case with SDSS DR12 BAO data. This provides another example of the feature noted in the literature that different supernova datasets can lead to different cosmological parameter estimates. It may be worth noting that some of these differences become less pronounced in the DES-Dovekie reanalysis \cite{DES:2025sig}.

\begin{figure}
\centering
\includegraphics[width=.6\textwidth]{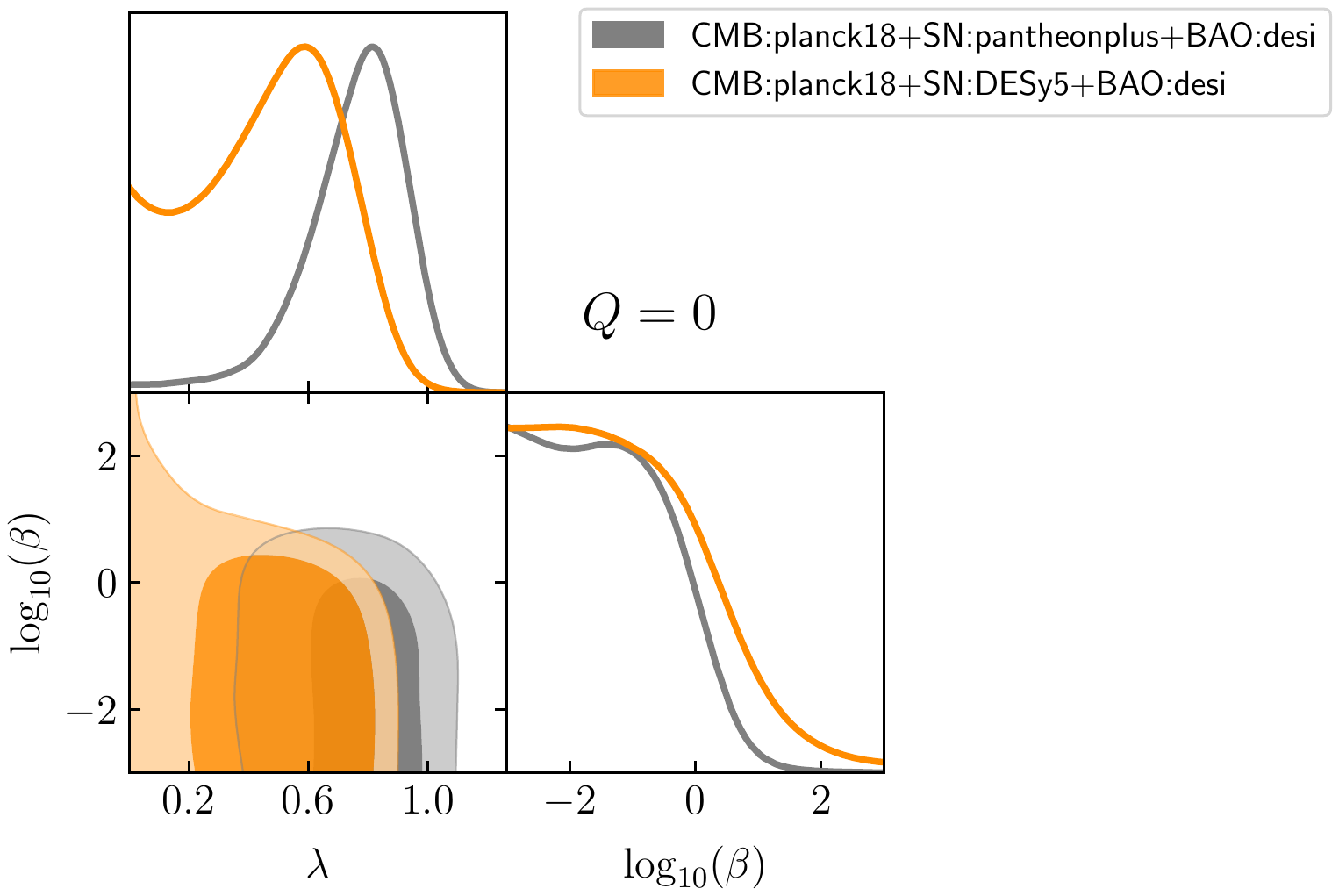}
\caption{Constraints on the model parameters $\log_{10}\beta$ and $\lambda$ for the pure momentum-transfer case ($Q=0$), obtained by combining DESI-DR2 BAO measurements with the two SN datasets DES-Y5 and Pantheon+.}
\label{paramsQ0}
\end{figure}

In order to compare our results with those previously reported in \cite{Pourtsidou2026}, and to better understand the impact of the selected datasets on the observational constraints of the model, we also perform an MCMC analysis for the pure momentum-transfer model ($Q=0$). We show the posterior in the $\lambda-\log_{10}\beta$ plane in Fig.~\ref{paramsQ0}, where we compare the effect of using 
Pantheon+ or DES-Y5 supernova datasets. We find that a detection of a non-vanishing parameter $\lambda$ becomes significant (at more than $2\sigma$) when using DES-Y5, while the posterior obtained with Pantheon+ extends to $\lambda=0$ within the $2\sigma$ region. Again, this highlights the importance of using different supernovae data for the detection of certain model parameters. However, let us emphasise once more that the DES results may change when using the latest DES-Dovekie reanalysis \cite{DES:2025sig}, which appeared upon completion of this work and brings them into closer agreement with the Pantheon+ results. Finally, we should note a certain disagreement between our results for the pure momentum case and those reported in \cite{Pourtsidou2026}. It was found in that work that a detection of the momentum transfer can be obtained by fixing the slope of the potential $\lambda$ to the lower bound conjectured from string theory, i.e., by fixing $\lambda=\sqrt{2}$. This value is at odds with our constraints, while it is within the $1\sigma$ region of the results obtained in \cite{Pourtsidou2026}. Furthermore, by fixing the slope of the potential $\lambda$ we do not obtain a detection of the momentum transfer, as it is clear from Fig. \ref{paramsQ0} where we can see that the noninteracting case is in the $1\sigma$ region of any slice of fixed $\lambda$. The only difference in our dataset in Fig. \ref{paramsQ0} and those used in \cite{Pourtsidou2026} is that we are not including Planck lensing likelihood in our analysis, but this alone cannot explain the discrepancy (as a matter of fact, we have explicitly checked this). We have not been able to identify the source of the discrepancy, which could be some underlying assumptions on the models.

\subsection{Constraints on the 
$\nu$-QuintC model}
\label{QuintCmnu}

\begin{figure}
\centering
\includegraphics[width=1.\textwidth]{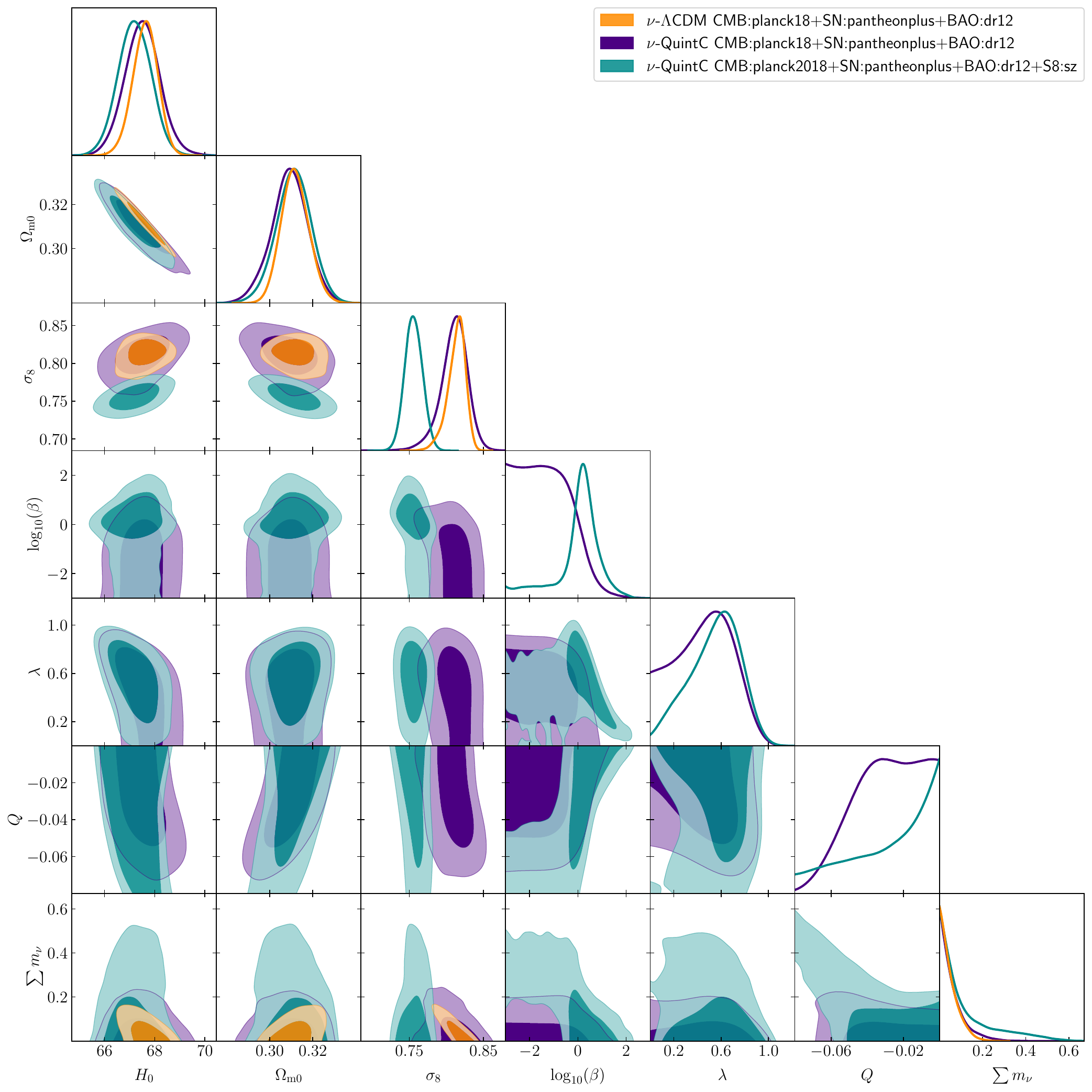}
\caption{Constraints on the $\nu$-QuintC model including a single massive neutrino with mass $m_\nu$.
}
\label{QuintCtrianglemnu}
\end{figure}

We finally address the main question of this paper, namely whether allowing the neutrino mass to vary can spoil the detection of the momentum transfer due to a potential degeneracy between $m_\nu$ and $\beta$. To this end, we consider the $\nu$-QuintC model described in Sec.~\ref{Sec:Modelanddata}, in which we include one massive neutrino with a varying mass $m_\nu$, and confront it with the Baseline I and Baseline I+$S_8$ datasets. The results of the MCMC analysis are shown in Fig.~\ref{QuintCtrianglemnu} and Table~\ref{tableMCMCbase}. 

By inspecting the posterior distribution of $\log_{10}\beta$ when including the $S_8$ data (green), we observe a pronounced peak around $\log_{10}\beta\simeq 0$, similar to that shown in Fig.~\ref{QuintCtriangleS8} for the case of a fixed neutrino mass. However, in Fig.~\ref{QuintCtrianglemnu} the posterior develops a flat tail extending towards $\beta\simeq 0$, such that the $\beta=0$ case lies within the $2\sigma$ region. We also find that the confidence regions in the planes involving $\beta$ no longer close at $2\sigma$ when the $S_8$ data (green) are included. 
Therefore, the detection of the momentum transfer is compromised when the neutrino mass is allowed to vary and is therefore not robust. This result contrasts with the findings of \cite{BeltranJimenez:2024lml}, where it was shown that allowing the neutrino mass to vary has only a marginal impact on the detection of a pure momentum transfer.

\begin{figure}
\centering
\includegraphics[width=0.45\textwidth]{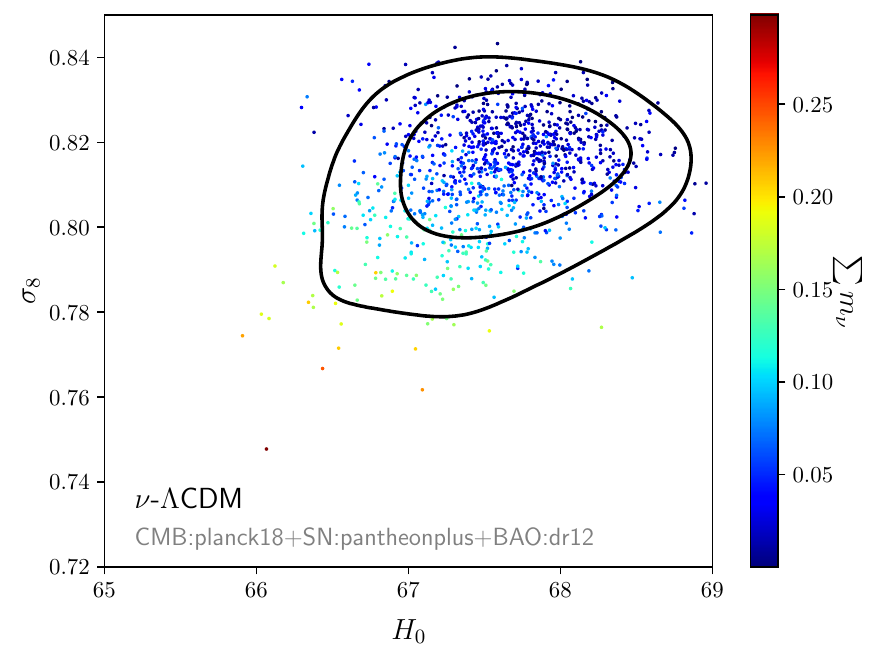}
\includegraphics[width=0.45\textwidth]{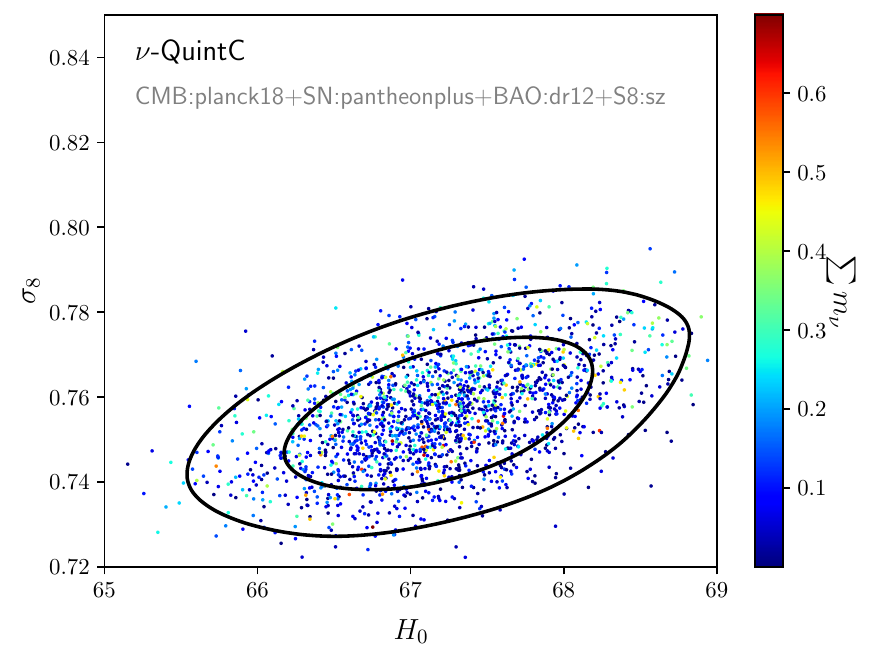} \\
\includegraphics[width=0.325\textwidth]{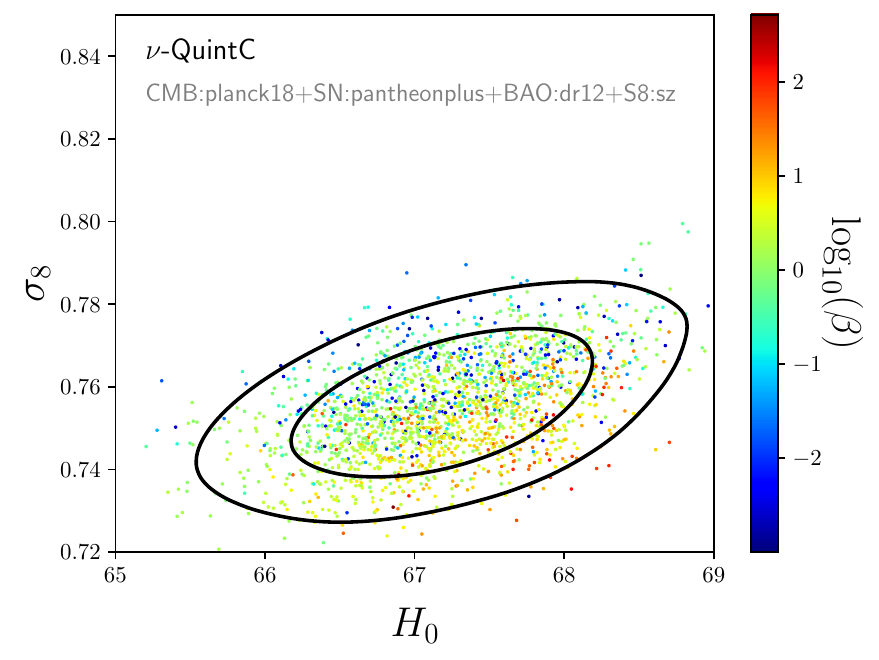}
\includegraphics[width=0.325\textwidth]{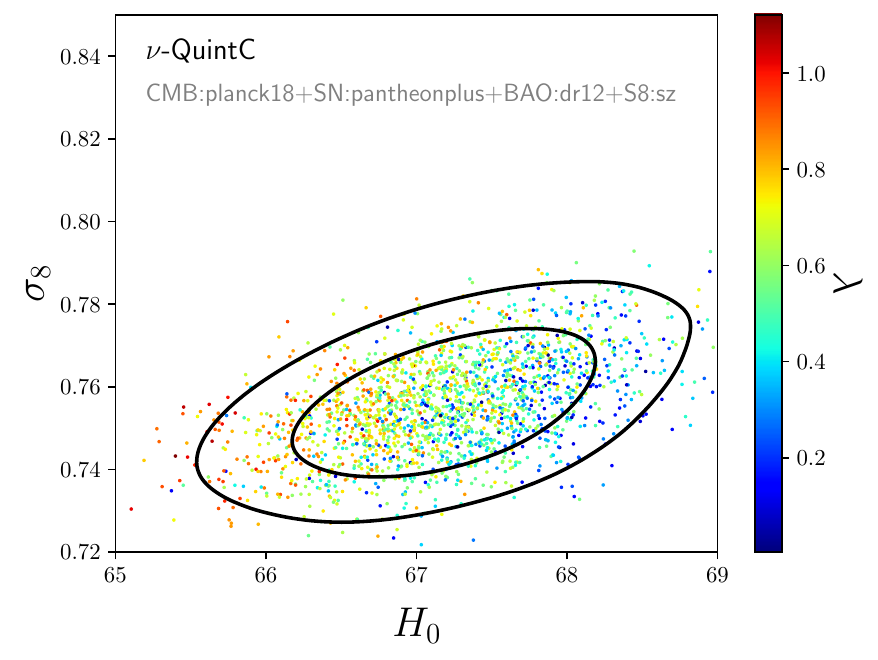}
\includegraphics[width=0.325\textwidth]{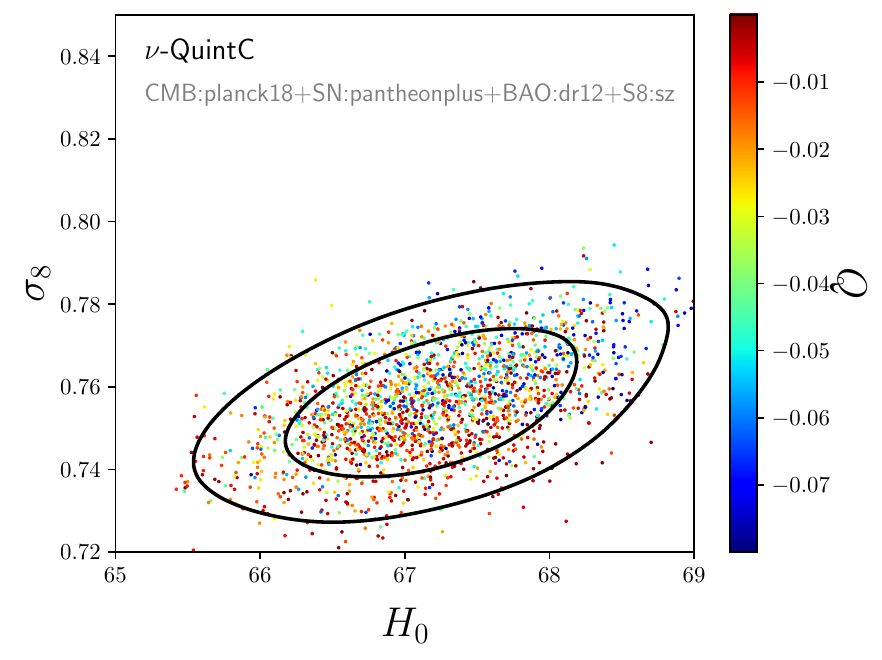}
\caption{Constraints on the $\nu$-$\Lambda$CDM and $\nu$-QuintC models in the $H_0$-$\sigma_8$ plane, using the Baseline I and Baseline I+$S_{8,{\rm SZ}}$ datasets, respectively. The points are coloured by the total neutrino mass $\sum m_\nu$ in the upper panel, and by the parameters $\lambda$, $\beta$, and $Q$ for the $\nu$-QuintC model in the lower panel.}
\label{H0sig8plane}
\end{figure}

\begin{figure}
\centering
\includegraphics[width=0.325\textwidth]{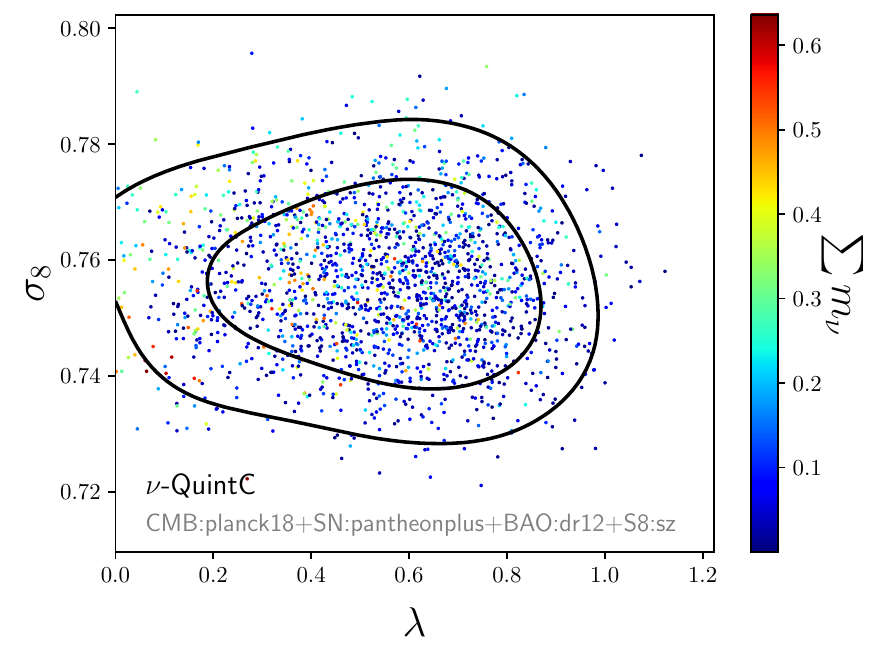}
\includegraphics[width=0.325\textwidth]{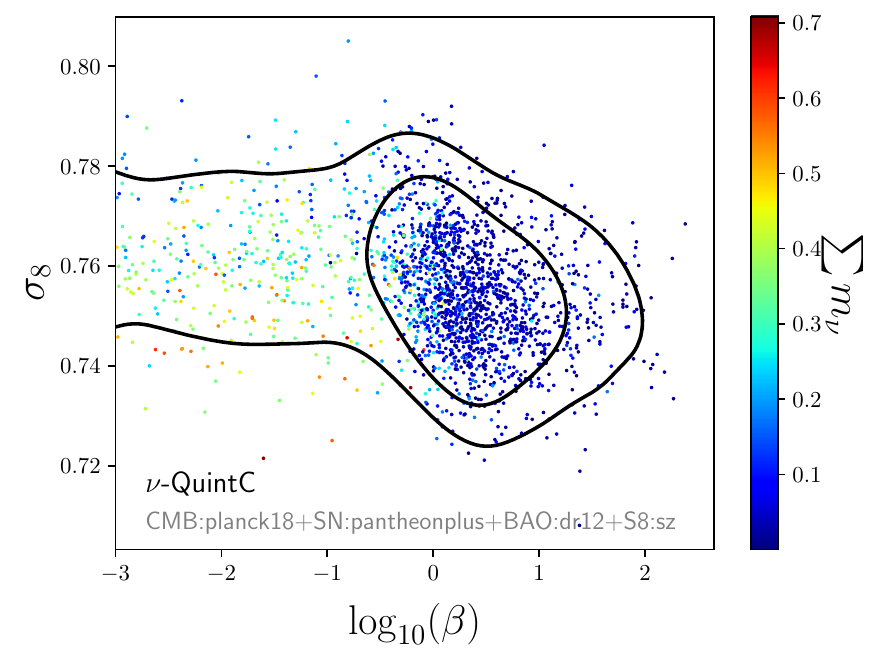}
\includegraphics[width=0.325\textwidth]{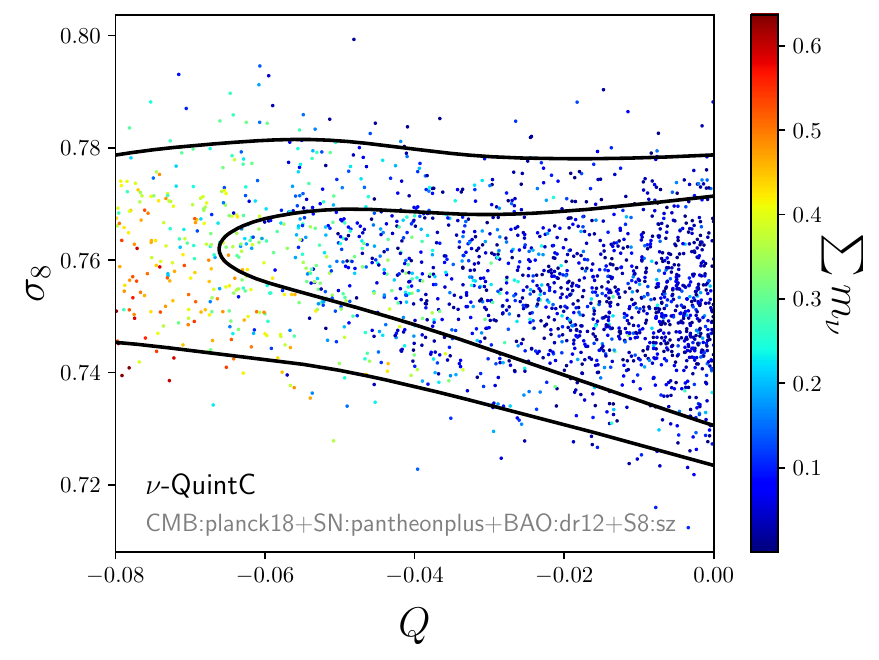}
\caption{Constraints on the value of $\sigma_8$ for the $\nu$-QuintC model, coloured by the total neutrino mass $\sum m_\nu$. 
These two-dimensional planes involving $\sigma_8$ show no degeneracy with the neutrino mass. }
\label{paramsVSs8VSmnu}
\end{figure}

It is interesting to note in Fig.~\ref{QuintCtrianglemnu} that the degradation of the momentum-transfer detection is not entirely driven by allowing the neutrino mass to vary, but rather by its combination with a non-vanishing energy exchange. 
This becomes apparent from the fact that the tail of the posterior extending towards $\beta \simeq 0$ corresponds to a region with 
non-vanishing $Q$, as can be seen 
in the $Q$--$\log_{10}\beta$ plane of Fig.~\ref{QuintCtrianglemnu}. 
Indeed, if we consider the $Q=0$ slice in that plane, the posterior for $\beta$ no longer extends to $\beta\simeq 0$ and closes at the $2\sigma$ level.

The combined effect of the energy transfer and the varying neutrino mass in weakening the momentum-transfer detection is further supported by the correlation observed in the tail of the posterior in the 
$Q$--$\Sigma m_\nu$ plane. 
We can therefore conclude that 
the impact of allowing the neutrino mass to vary in scenarios featuring momentum transfer must be assessed on a model-by-model basis. In general, no universal statement can be made, since additional degenerate directions with other parameters may arise.

In Fig.~\ref{H0sig8plane} we show how the posteriors in the $\sigma_8$--$H_0$ plane correlate with different parameters of the interacting model. In the upper left 
panel we show this plane for $\Lambda$CDM with a varying neutrino mass, illustrating the well-known anti-correlation between $\sigma_8$ and the neutrino mass. In this case, lower values of $\sigma_8$ correspond to larger values of $m_\nu$, which in turn tend to drive the posterior towards smaller values of $H_0$. This reflects the well-known issue that alleviating one tension tends to exacerbate the other. 

In the upper right panel we show the same plane for the $\nu$-QuintC model. In this case, the correlation with the neutrino mass disappears because the value of $\sigma_8$ can be substantially suppressed by the momentum transfer governed by $\beta$, without affecting the background evolution. Despite the disappearance of this correlation with the neutrino mass, we still observe a correlation between $\sigma_8$ and $H_0$, such that smaller values of $\sigma_8$ tend to prefer smaller values of $H_0$. However, it is important to note that this correlation does not imply that alleviating the $\sigma_8$ tension necessarily worsens the Hubble tension, since the momentum transfer mainly shifts the posterior towards lower values of $\sigma_8$ without significantly affecting the posterior of $H_0$.

In the lower panels of Fig.~\ref{H0sig8plane}, 
we show the posterior distribution in the 
$\sigma_8$--$H_0$ plane and its dependence on the interacting quintessence parameters. In the lower-left panel, we see that the momentum transfer is clearly responsible for producing smaller  values of $\sigma_8$. 
The middle panel shows that the slope parameter $\lambda$ correlates with $H_0$, such that smaller values of $\lambda$ correspond to higher values of the Hubble constant. Finally, the lower-right panel shows that the energy-exchange parameter $Q$ does not play any significant role in the tensions, as it appears to be completely uncorrelated 
with the $\sigma_8$--$H_0$ plane. However, as pointed out above, this parameter is crucial in spoiling the detection of the momentum transfer when the neutrino mass is allowed to vary. This behaviour is also evident in Fig.~\ref{paramsVSs8VSmnu}, where the $\sigma_8$ plane for each model parameter is shown, colour-coded by $m_\nu$. 
We observe that, while variations in $m_\nu$ play no significant role in the cases of $\beta$ and $\lambda$, the parameter $Q$ exhibits a clear gradient correlated with $m_\nu$.

Finally, we compare the coupled quintessence model with $\Lambda$CDM by evaluating $\Delta \chi^2$ and $\Delta{\rm AIC}$ \cite{Akaike1974}. Without including the $S_8$ measurement, the improvement of the QuintC model over $\Lambda$CDM is negligible ($\Delta \chi^2 = 0.96$ and $\Delta{\rm AIC}=5.04$ for three additional parameters). However, once $S_{8,{\rm SZ}}$ is included, the QuintC model provides a substantially better fit ($\Delta \chi^2 = 17.18$ and $\Delta{\rm AIC}=23.18$), corresponding to a strong preference according to the standard AIC criterion. When the additional parameter $m_{\nu}$ is allowed to vary in both models, this preference is significantly reduced ($\Delta \chi^2 = 8.61$ and $\Delta{\rm AIC}=2.61$ including $S_{8,{\rm SZ}}$). However, it is important to note that $\Lambda$CDM will increase the Hubble tension, while the QuintC model will remain at the same level for the reasons explained above. Overall, the statistical preference for the QuintC model is mainly driven by the inclusion of the $S_{8,{\rm SZ}}$ measurement.

\section{Conclusions}
\label{results}

In this paper, we have placed observational constraints on the interacting model of DE and DM 
described by the action (\ref{action}). The model is characterized by two constants, $Q$ and $\beta$, which mediate energy and momentum transfer, respectively, 
together with the slope $\lambda$ of the exponential potential. 
Unlike the MCMC analysis performed 
in \cite{Liu:2023mwx}, we have taken into account the $S_8$ data and have also allowed the mass of a neutrino to vary. We have used not only the Pantheon SNe~Ia and Planck2018 CMB data, but also the BAO data from DESI DR2 and the SNe~Ia data from DES-Y5.

The previous analysis of Ref.~\cite{Liu:2023mwx}, performed without varying the neutrino mass, reported evidence for a nonvanishing momentum-transfer parameter $\beta$ at the $2\sigma$ confidence level, even in the absence of the $S_8$ data. 
Repeating the analysis with the same dataset (Baseline~I), we instead find only an upper bound on $\beta$, with no detection of momentum transfer at the $2\sigma$ level. We have traced this discrepancy to an improper sampling of the momentum-transfer parameter introduced in \cite{Liu:2023mwx}. When the sampling is performed correctly, the results become fully consistent with those obtained in this work. Including the $S_8$ data from Planck SZ measurements, and without varying the neutrino mass, we find $2\sigma$ evidence for momentum transfer. When the $S_8$ measurement from DES-Y3 is used instead, however, no significant evidence for a nonvanishing $\beta$ is found. This highlights the crucial importance of the measurement of $S_8$ employed for the scenario under consideration in this work, which does not align with previous findings in the literature where the constraints were relatively robust to the value of $S_8$.

Allowing the neutrino mass to vary, we find that it can spoil the detection of momentum transfer. Interestingly, this effect arises not from the neutrino mass alone, but from the combined effect of the neutrino mass and the energy exchange present in these models, which turn out to be correlated. 
Our findings highlight the importance of accounting for massive neutrinos when assessing the significance of a momentum-transfer detection from low-redshift probes. This is not only because massive neutrinos suppress the growth of structures, similarly to the effect of momentum transfer, but also because they can introduce new degeneracies with other model parameters, such as the energy exchange present in the coupled quintessence scenario considered here. 

By comparing $\chi^2$ and AIC between the QuintC model and $\Lambda$CDM, we find that the former is strongly favored when the $S_{8,{\rm SZ}}$ data are included in the analysis. However, this preference for the coupled quintessence model over $\Lambda$CDM is weakened when the neutrino mass is allowed to vary, due to the degeneracy with the coupling $Q$ mentioned above. However, the Hubble tension will worsen for $\Lambda$CDM, but not for the QuintC model. Therefore, the robustness of the detection of momentum transfer depends on the underlying model and on the treatment of the neutrino mass.

\section*{Acknowledgements}

JBJ and FATP thank the hospitality of the Department of Physics at Waseda University where this work was initiated and the Institute of Theoretical Astrophysics of the University of Oslo where part of this work was performed. ST thanks the members of the University of Salamanca for their warm hospitality during his stay. JBJ and FATP acknowledge the support from grants PID2024-158938NB-I00 funded by MICIU/AEI/10.13039/ 501100011033 and by “ERDF A way of making Europe” and the Project SA097P24 funded by Junta de Castilla y Le\'on. FATP acknowledges the support of {\it Programa Propio I+D+i (2024)}, by Universidad Politécnica de Madrid, and {\it Programa de estancias de movilidad,  Modalidad Junior José Castillejo (2023)} by Ministerio de Ciencia, Innovación y Universidades, Spain.
ST thanks JSPS KAKENHI Grant No.~22K03642 and Waseda University Special Research Projects (Nos.~2025C-488 
and 2025R-028) for their support. XL is supported by the I+D grant PID2023-149018NB-C42 and the Grant IFT Centro de Excelencia Severo Ochoa No CEX2020-001007-S, funded by MCIN/AEI/10.13039/501100011033.
KI is supported by the JSPS grant numbers 21H04467 and 24K00625, and the JST FOREST Program JPMJFR20352935.

\bibliography{bib}

\end{document}